\documentclass[sigconf, nonacm]{acmart}
\AtBeginDocument{%
 \providecommand\BibTeX{{%
 \normalfont B\kern-0.5em{\scshape i\kern-0.25em b}\kern-0.8em\TeX}}}

\usepackage[utf8]{inputenc}
\setcopyright{acmcopyright}

\acmDOI{10.475}

\newcommand\vldbdoi{XX.XX/XXX.XX}
\newcommand\vldbpages{XXX-XXX}
\newcommand\vldbvolume{19}
\newcommand\vldbissue{X}
\newcommand\vldbyear{2026}
\newcommand\vldbauthors{\authors}

\usepackage{amsthm}
\usepackage{url} 
\usepackage{epstopdf}

\usepackage[linesnumbered,ruled,lined,boxed,noend]{algorithm2e}
\newcommand{\myparagraph}[1]{\vspace{0.5\baselineskip}\noindent{\textbf{#1.}}~}
\usepackage{xspace}

\SetKwComment{Comment}{$\triangleright$\ }{}

\newtheorem{example1}{Example}

\newtheorem{definition1}{Definition}

\newtheorem{lemma1}{Lemma}

\usepackage{subfigure}
\makeatletter
\renewcommand{\@thesubfigure}{\hskip\subfiglabelskip}
\makeatother

\usepackage{epsfig}
\usepackage{epstopdf}

\usepackage{enumitem}
\usepackage{titlesec}

\usepackage{color,soul}
\RequirePackage[prologue]{xcolor}
\definecolor[named]{ACMPurple}{cmyk}{0.55,1,0,0.15}
\definecolor{Maroon}{cmyk}{0, 0.87, 0.68, 0.32}
\definecolor{Brown}{rgb}{0.59, 0.29, 0.0}
\definecolor{Green}{rgb}{0,1,0}
\definecolor{NavyBlue}{rgb}{0.0, 0.0, 0.5}

\setul{0.5ex}{0.3ex}
\definecolor{Green}{rgb}{0,1,0}
\setulcolor{Green}

\definecolor{issuecolor}{RGB}{0,166,81}

\newcounter{cN}
\usepackage{etoolbox}
\usepackage{marginnote}
\makeatletter
\let\oldmarginnote\marginnote
\renewcommand*{\marginnote}[1]{%
	\begingroup%
	\ifodd\value{page}
	\if@firstcolumn\normalmarginpar\fi
	\else
	\if@firstcolumn\else\normalmarginpar\fi
	\fi
\textbf{}	\oldmarginnote{\textcolor{Brown}{#1}}%
		\oldmarginnote{}
	\endgroup%
}
\makeatother
\usepackage{soul}
\usepackage{etoolbox}

\usepackage{balance}
\usepackage{tablefootnote}
\usepackage{booktabs} 
\usepackage{tikz}  

\newcommand{\spadas}{{{\textsf{Spadas}}}\xspace}
\newcommand{\IA}{{{\textsf{IA}}}\xspace}
\newcommand{\GBO}{{{\textsf{GBO}}}\xspace}
\newcommand{\Haus}{{{\textsf{Haus}}}\xspace}
\newcommand{\ExactHaus}{{{\textsf{ExactHaus}}}\xspace}
\newcommand{\ApproHaus}{{{\textsf{ApproHaus}}}\xspace}
\newcommand{\RangeP}{{{\textsf{RangeP}}}\xspace}
\newcommand{\NNP}{{{\textsf{NNP}}}\xspace}
\newcommand{\RangeS}{{{\textsf{RangeS}}}\xspace}
\newcommand{\ScanHaus}{{{\textsf{ScanHaus}}}\xspace}
\newcommand{\ScanGBO}{{{\textsf{ScanGBO}}}\xspace}
\newcommand{\Origin}{{{\textsf{Origin}}}\xspace}
\newcommand{\IncHaus}{{{\textsf{IncHaus}}}\xspace}
\newcommand{\INNE}{{{\textsf{INNE}}}\xspace}
\newcommand{\KNN}{{{\textsf{kNN}}}\xspace}

\usepackage{makecell}

\usepackage{orcidlink}

\hypersetup{pdfstartpage=17}  

\setcopyright{acmcopyright}
\copyrightyear{2026}
\acmYear{2026}
\acmDOI{XXXXXXX.XXXXXXX}

\begin{document}
\setlength{\textfloatsep}{0cm}

\title{A Unified Approach for Multi-Granularity Search over Spatial Datasets
}

\author{Wenzhe Yang}
\affiliation{%
  \institution{Hohai University}
}
\email{wenzheyang@hhu.edu.cn}

\author{Sheng Wang$^{\ast}$}
\affiliation{%
  \institution{Wuhan University}
}
\email{swangcs@whu.edu.cn}

\author{Shixun Huang}
\affiliation{%
  \institution{The University of Wollongong}
}
\email{shixun_huang@uow.edu.au}

\author{Hao Liu}
\affiliation{%
  \institution{Wuhan University}
}
\email{liu_hao@whu.edu.cn}

\author{Yuan Sun}
\affiliation{%
  \institution{La Trobe University}
}
\email{yuan.sun@latrobe.edu.au}

\author{Juliana Freire}
\affiliation{%
  \institution{New York University}
}
\email{juliana.freire@nyu.edu}

\author{Zhiyong Peng}
\affiliation{%
  \institution{Wuhan University}
}
\email{peng@whu.edu.cn}

\begin{abstract}
There has been increased interest in data search as a means to find relevant datasets or data points in data lakes and repositories. Although approaches have been proposed to support spatial dataset search and data point search, they consider the two types of searches independently.
To enable search operations ranging from the coarse-grained dataset level to the fine-grained data point level, we provide an integrated one that supports diverse query types and distance metrics. In this paper, we focus on designing a multi-granularity  \underline{\textbf{spa}}tial \underline{\textbf{da}}ta \underline{\textbf{s}}earch system, called \spadas, that supports both dataset and data point search operations. To address the challenges of the high cost of indexing and susceptibility to outliers, we propose a unified index that can drastically improve query efficiency in various scenarios by organizing data reasonably and removing outliers in datasets. Moreover, to accelerate all data search operations, we propose a set of pruning mechanisms based on the unified index, including fast bound estimation, approximation technique with error bound, and pruning in batch techniques, to effectively filter out non-relevant datasets and points. Finally, we report the results of a detailed experimental evaluation using six spatial data repositories, achieving orders of magnitude faster than the state-of-the-art algorithms and demonstrating the effectiveness by case study. An online spatial data search system of \spadas is also implemented and made accessible to users. 
\end{abstract}
\maketitle
\begingroup\small\noindent\raggedright\textbf{PVLDB Reference Format:}\\
\vldbauthors. A Unified Approach for Multi- Granularity Search over Spatial Datasets. PVLDB, \vldbvolume(\vldbissue): \vldbpages, \vldbyear.\\
\href{https://doi.org/\vldbdoi}{doi:\vldbdoi}
\endgroup
\begingroup
\renewcommand\thefootnote{}\footnote{\noindent
This work is licensed under the Creative Commons BY-NC-ND 4.0 International License. Visit \url{https://creativecommons.org/licenses/by-nc-nd/4.0/} to view a copy of this license. For™ any use beyond those covered by this license, obtain permission by emailing \href{mailto:info@vldb.org}{info@vldb.org}. Copyright is held by the owner/author(s). Publication rights licensed to the VLDB Endowment. \\
\raggedright Proceedings of the VLDB Endowment, Vol. \vldbvolume, No. \vldbissue\ %
ISSN 2150-8097. \\
\href{https://doi.org/\vldbdoi}{doi:\vldbdoi} \\
}\addtocounter{footnote}{-1}\endgroup

\vspace{-1em}

\section{Introduction}
\label{sec:intro}
\begin{sloppypar}



The development of sensor technology, GPS-enabled mobile devices, and wireless communication brings prosperity to spatial data, and the emergence of a large number of spatial data has fueled the demand for searching spatial data~\cite{Degbelo2019,Bouros2019,Nandan2014,Zacharatou2019,Wang2019d,YangWZ2022,DeFernandesVasconcelos2017}. Since a spatial dataset usually consists of a set of spatial data points~\cite{YangWZ2022}, various spatial \emph{dataset} search~\cite{YangWZ2022,Li0SWYY24,LiPY24} and \emph{data point} search systems~\cite{foursquare,ChenCCT15,Dong2019b} have been developed. For example, Foursquare~\cite{foursquare} helps users find the best location points in any city around the world. OpenStreetMap~\cite{openStreetMap} provides search functionality over a vast collection of spatial datasets contributed by users worldwide. However, these systems primarily support keyword-based searches limited to published dataset metadata~\cite{Chirigati2021, ZhongLZ23,ChenLS2020}.

The published spatial metadata is often missing or incomplete~\cite{auctus}, rendering keyword-based queries insufficient for helping users locate datasets or data points. Therefore, various query operations that rely on raw data have been developed to identify spatial datasets and data points. 
For example, existing spatial dataset search engines treat a dataset as the smallest return unit.  In contrast, in the existing spatial data point search engines, a data point is the smallest return unit. In what follows, we use a traffic analysis example to introduce these two types of searches.

\begin{figure}
\setlength{\abovecaptionskip}{0 cm}
\setlength{\belowcaptionskip}{0 cm}
	\centering
	\includegraphics[width=0.95\linewidth]{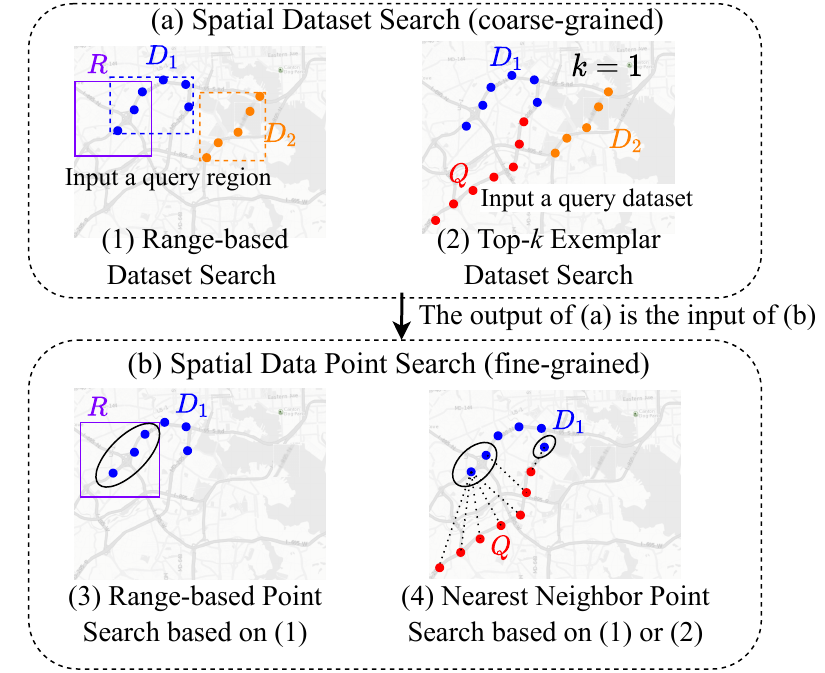}
	\caption{{The illustration of our multi-granularity search research over spatial datasets, which seamlessly integrates the dataset search (including (1) and (2) based on three distance metrics) and data point search (including (3) and (4)). 
  }}
	\label{fig:intro1}
\end{figure}
\vspace{-0.1cm}
\begin{example1}
Consider a scenario where traffic analysts want to analyze traffic patterns in a metropolitan area. Firstly, they can define a query range of $R$~\cite{auctus,Zacharatou2019,BrinkhoffKS93} to locate all relevant traffic datasets within that area for analysis, as illustrated on the left side of Figure\ref{fig:intro1}(a). Then, to identify the most efficient commuting routes, the analysts can input a specific traffic dataset $Q$ and search for datasets similar to $Q$ based on defined similarity measures~\cite{YangWZ2022,Wang2017,Wangsheng2018} for data analysis, illustrated on the right side of Figure~\ref{fig:intro1}(a). This search manner is known as exemplar search~\cite{Rezig2021,YewLML23, WangMGBCDSW24,Li0SWYY24,Mottin2014a,YangZ0H0020}.
Furthermore, if the analysts want to find points of interest, such as gas stations or rest areas along the searched routes $D$, they can specify a query range $R$ along with the relevant traffic dataset to retrieve these points, as depicted on the left side of Figure\ref{fig:intro1}(b)\cite{foursquare,Wang2019d,ZHANGSongnian2021,Qi2020}. Additionally, on the right side of Figure\ref{fig:intro1}(b), the analysts can identify the nearest gas stations for each location in the dataset $Q$~\cite{Qi2020,ZHANGSongnian2021,Dong2019b,ChoudhuryCSC16,CorralMTV2000}, which helps commuters find convenient stops along their routes~\cite{YinWWCZ17,yao2016poi}.
%


\end{example1}
\noindent\textbf{What is the Problem?} 
Despite the efforts made in the field of spatial data search systems, existing systems focus mainly on one type of search. There is currently no system that has successfully integrated all types of searches into a unified search framework. In addition, the existing systems, such as Auctus~\cite{auctus} and OpenStreetMap~\cite{openStreetMap}, are relatively simple in spatial data search and do not take into account the similarity measurement of spatial location distribution characteristics. 
Our study, depicted in Figure~\ref{fig:intro1}, considers all searches at both the coarse-grained spatial dataset level and the more fine-grained data point level. In spatial dataset search, we integrate range-based dataset search and top-$k$ exemplar dataset search based on several distance metrics.




\begin{figure}
\setlength{\abovecaptionskip}{-0.1 cm}
\setlength{\belowcaptionskip}{0 cm}
	\centering
	\includegraphics[width=1.0\linewidth]{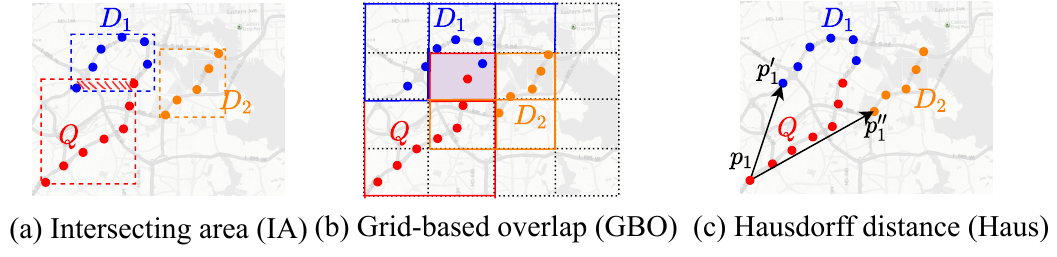}
	\caption{The illustration of three distance metrics, where (a) shows that the IA of $Q$ and $D_1$ is the size of their minimum bounding rectangles' overlapping areas, i.e., the rectangle filled with red lines, (b) shows that the GBO of $Q$ and $D_1$ is 1 since there is one overlapping cell between $Q$ and $D_1$, and (c) shows that the Haus of $Q$ and $D_1$ is the maximum nearest neighbor distance, i.e., the distance from the point $p_1$ of $Q$ to the point $p_1'$ of $D_1$.}
\label{fig:motivation}
\end{figure}

 As shown in Figure~\ref{fig:motivation}, the intersecting area (\IA)~\cite{Larson2011,Degbelo2019}  is useful when assessing the overlap between public transportation routes, which can help identify congestion points to optimize traffic management.
Compared to \IA, grid-based overlap (\GBO)~\cite{Nandan2014} can capture the density of traffic in the urban grid in a finer granularity and optimize traffic routes.
In addition, calculating the Hausdorff distance (\Haus)~\cite{Taha2015,Nutanong2011,Frontiera2008} can help analysts understand the accessibility of bus stops for residents (the formal definition is in Section~\ref{sec:distanceMetrics}). 
Moreover, building upon the outcomes of the dataset search, we further enhance our capabilities to facilitate a finer-grained data point search process, wherein the results obtained from the dataset search are fed to the subsequent data point search.


%

\noindent\textbf{What are the Challenges?}
Our goal is to design a multi-granularity spatial data search system that seamlessly supports coarse-grained dataset search based on multiple distance metrics and fine-grained data point search. There are several challenges: Firstly, A na\"{i}ve way is to build separate indexes for each query type. However, this method incurs significant costs in both storage and index construction time. Furthermore, the presence of outliers in spatial datasets, such as erroneous data points caused by GPS errors, can severely affect the accuracy and efficiency of similarity computation~\cite{Wang2012}. Finally, although the index can quickly locate data matching query criteria, it does not effectively reduce the number of candidate points to examine. Thus, there is an urgent need for efficient acceleration strategies to speed up a rich set of searches.

\noindent\textbf{How the Challenges are Addressed?}
To address the challenges above, we develop a multi-granularity spatial data search system, called \spadas. Specifically, we design a \textit{unified index} \cite{Wang2018a,Hoang-Vu2016} that can effectively organize both datasets and data points, enabling efficient pruning for various search scenarios. In addition, to further enhance system performance, 
we propose a set of novel and highly efficient \textit{pruning mechanisms}, such as efficient bound estimation or error-bounded approximation techniques, ensuring that irrelevant candidates are filtered early in the process, thereby reducing computational overhead and improving overall system performance. Our main contributions are summarized as follows.

\begin{itemize}[leftmargin=*]
    \item 
    We propose a multi-granularity spatial data search system, called \spadas, to efficiently support coarse-grained dataset search based on multiple distance metrics and fine-grained data point search (see Section \ref{sec:framework}). We also implement an online \spadas system\footnote{\url{http://swangdbs.com/prototype/spadas/}} that can be accessed by users.

    \item To support both searches with different query types and distance metrics, we specifically design a unified index to enhance the efficiency of data search operations by efficiently organizing spatial data, while mitigating the influence of outliers on the accuracy of similarity measures between datasets (see Section \ref{sec:index}).
        
    \item To accelerate multiple types of data search operations, we propose a range of pruning mechanisms over the unified index, including tighter lower and upper bounds, approximation technique with error bound, and batch pruning, that effectively filter candidates and thus significantly reduce computational overhead (see Section \ref{sec:accelerating}). 
    

    \item  We conduct extensive experiments on six real-world data repositories, and the results show that our proposed \spadas can achieve orders of magnitude speedup than the state-of-the-art algorithms in dataset search and data point search. The case studies validate the effectiveness of our search system (see Section \ref{sec:exp}).



\end{itemize}

 \end{sloppypar}


\section{Related Work}
\label{sec:liter}
\begin{sloppypar}
{
In this section, we give a literature review on spatial dataset search and spatial data point search. 
}
\vspace{-0.5em}
\subsection{{Spatial Dataset Search}}
\vspace{-0.5em}
%



%

\myparagraph{Overlap-based Similarity Search}
For a user with a query dataset, the overlap-based similarity search ranks candidate datasets based on the size of their overlap area. Computing the overlapping area between two Minimum Bounding Rectangles (MBRs) is a commonly used and straightforward measure. Frontiera et al. \cite{Frontiera2008} proposed several variants by considering the query and the dataset's range. 
Vasconcelos et al. \cite{DeFernandesVasconcelos2017} and Degbelo et al.~\cite{Degbelo2019} applied it as the main ranking metric for an open government dataset search portal.
{More recently, with the rise of grid partition technology \cite{Zacharatou2019,Winter2019}, a more fine-grained similarity measure based on the number of overlapping grid cells began to appear}, i.e., modeling each dataset as a grid with points inside by dividing the space equally and using the intersection of the grids as the similarity. 

\myparagraph{Point Set Similarity Search}
Point set similarity search mainly calculates the distance between each pair of points to assess the similarity between two datasets. For instance, Hausdorff distance~\cite{Adelfio2011,Nutanong2011,Degbelo2019,Gao2014,Chen2017g,Li2017e,Taha2015} measures the similarity between two datasets by calculating the maximum distance from any point in one dataset to its nearest point in the other dataset.
Specifically, Nutanong et al.~\cite{Nutanong2011} derived the lower and upper bounds of Hausdorff and combined them with the branch-and-bound strategy to improve efficiency. Degbelo et al.~\cite{Degbelo2019} conducted experiments to verify that the Hausdorff distance has higher precision than the overlapped area in dataset search. However, the complexity of the existing upper and lower bounds is still high, and we need more efficient bound estimation to speed up the Hausdorff computation. In addition, earth mover's distance~\cite{YangWZ2022, Li0SWYY24, LiPY24} is a more fine-grained point set similarity measure compared to Hausdorff. However, it is a more expensive distance as it has a cubic time complexity.



\vspace{-0.3cm}
\subsection{{Spatial Data Point Search}}



\myparagraph{Range-based search} 
As one of the fundamental operators in spatial databases, the goal of range-based search is to find all
objects falling within the query region. For example, Hariharan et al.~\cite{HariharanHLM07} constructs a location-based search engine, which supports users to specify a query range, and then all points that fall into the query region are returned to the users. Zacharatou et al.~\cite{Zacharatou2019} proposed a spatial bundled dataset search over different domains using spatial ranges. In the spatial dataset search system Auctus~\cite{Chirigati2021,auctus}, range queries are supported by returning spatial datasets that overlap with the user-specified query region. Papadopoulos et al.~\cite{PapadopoulosM98} also studied multiple-range query optimization strategies to improve search efficiency by grouping multiple queries.

\myparagraph{$k$ Nearest Neighbor search}
The $k$NN~\cite{YuCWS07,ChenCCT15,Qi2020,G2010,NutanongS13} is defined as finding $k$ points in a given set that are closest to a given query point. The $k$NN queries have been extensively studied in the literature. To find the $k$ closest neighbor for any given query point, we can utilize the fundamental $k$NN search approach, which is often called the brute force (BF) method. For the given query points, it scans the entire dataset to find the $k$ closest points based on the distances between the query point and all other data points, which is computationally intensive. Thus, to efficiently accelerate the $k$NN search, a variety of indexes with different heuristic improvements have been proposed, such as R tree~\cite{Guttman84}, K-Dimensional tree (KD-tree)~\cite{Sproull91}, and Ball-tree~\cite{Omohundro1989}. While these indexes can be implemented to find $k$ nearest neighbor points for a given point, they cannot support both spatial dataset search and spatial data point search, which is the focus of this work. 

\myparagraph{\underline{Remarks}} 
No work has integrated the above spatial dataset search and data point search. Although there exist some index techniques to speed up the search process, different types of queries usually require a dedicated index structure. They do not consider building a unified index, not to mention pruning mechanisms with the index to accelerate query processing.

\end{sloppypar}
\section{Preliminaries}
\label{sec:defi}


In this section, we first introduce the data model and multiple similarity metrics between spatial datasets. Subsequently, we introduce coarse-grained dataset search and fine-grained data point search, followed by the formalization of the problem of multi-granularity spatial data search. Frequently used notations are summarized in Appendix~\ref{appendix:notations}.

\vspace{-0.5em}
\subsection{Data Model}
\label{subsec:dataModelling}
\begin{definition1}{\textbf{(Spatial Data Point).}}
	\label{def:point}
	A spatial data point $p=(x, y, v_1, \cdots,$ $v_{d-2})$ is a d-dimension vector ($d\ge2$), which has at least a latitude $x$ and a longitude $y$, and can have $d-2$ other attributes.
\end{definition1}

For ease of presentation, we use 2-dimensional points for illustration in the subsequent descriptions. However, all search operations we support can be easily extended to multi-dimensional datasets~\cite{patro2015normalization}, as shown in Section~\ref{sec:exp}. . 

%

\begin{definition1}{\textbf{(Spatial Dataset).}}
	\label{def:dataset}
	A spatial dataset $D$ contains a set of $2$-dimension points, $i.e., D= \{p_1, p_2, \cdots, p_{|D|}\}$, where $|D|$ is the size of $D$. The minimum bounding rectangle (MBR) of $D$, whose sides are parallel to the $x$ and $y$ axes and minimally enclose all points, is specified by the bottom left corner $b^\downarrow$ and upper right corner $b^\uparrow$, i.e., 
 \begin{equation}
 \small
 D.b^\uparrow[i]=\max_{p \in D}p[i], \quad D.b^\downarrow[i]=\min_{p \in D}p[i], wrt., i\in\{1,2\}
 \end{equation}

\end{definition1}

\begin{definition1}{\textbf{(Spatial Data Repository).}}
	\label{def:datalake}
	A spatial data repository $\mathcal{D}$ is composed of a set of spatial datasets, $i.e., \mathcal{D} = \{D_1, \cdots, D_{|\mathcal{D}|}\}$, where $|\mathcal{D}|$ is the number of datasets in $\mathcal{D}$. 
 
 
\end{definition1}

When measuring the similarity between two datasets, existing systems~\cite{Degbelo2019,DeFernandesVasconcelos2017} mainly use the intersecting area between the MBR of datasets to measure the overlap ratio. 
To enable a more fine-grained measure of overlap, we introduce a structure called \textit{z-order signature} to represent each spatial dataset. 
Specifically, given an integer $\theta$, we partition the entire space containing the spatial data repository into a grid $\mathcal{C}_{\theta}$ of $2^\theta\times2^\theta$ equally-sized cells~\cite{Zacharatou2017,Cao2021a,QiaoHZWGW20}, where $\theta$ is referred to as the \textit{resolution}.

Each cell, denoted as $c$, can be treated as an element. 
The coordinates of cell $c$ can be converted into a unique non-negative integer $c.id$ by interleaving the binary representations of its coordinate values~\cite{Peng2016,Zacharatou2019,SieranojaF18,YangWZ2022}. This transformation yields consecutive IDs in the range $[0,2^{\theta} \times 2^{\theta}-1]$. In this way, a spatial dataset $D$ can be represented as a sorted integer set consisting of a sequence of cell IDs. The formal definition of \textit{z-order signature} is as follows.





\begin{definition1}{\textbf{($z$-order Signature).}}
\label{def:zorder}
Given a grid $\mathcal{C}_{\theta}$, the $z$-order signature of the spatial dataset $D$ is a sorted integer set consisting of a sequence of cell IDs, i.e., $z(D) = \{ c.id_1, \dots, c.id_{|z(D)|}\}$, where each element $c.id$ indicates that at least one spatial point from $D$ falls within the cell $c$.

\end{definition1}

\vspace{-0.3cm}
\subsection{Similarity Measures}
\label{sec:distanceMetrics}
To handle search scenarios with varying levels of granularity, we employ two types of measures to compute the dataset similarity. 

\subsubsection{\textbf{Overlap-based Similarity Measures}}
\label{sec:overlap-similarity}
\begin{definition1}\textbf{(Intersecting Area (IA))}
	\label{def:ia}
	Given a query dataset $Q$ and a spatial dataset $D$, the IA between $Q$ and $D$ is the size of the intersecting area of two given datasets' MBR, i.e., {IA}$(Q, D) = \prod_{i=1}^{2}l(i, Q, D), $
where $l(i, Q, D)$ denotes the intersecting length of two MBRs in $i$-th dimension~\cite{Zomorodian2000}. 

\end{definition1}




\begin{definition1}\textbf{(Grid-Based Overlap (GBO)).}
	\label{def:gbo}
 {Given a query dataset $Q$ and a spatial dataset $D$, the GBO between $Q$ and $D$ is the number of set intersections of two given datasets' z-order signature $z(Q)$ and $z(D)$, i.e., ${G}(Q, D)= |z(Q)\cap z(D)|$.}

\end{definition1}


\subsubsection{\textbf{Point-wise Similarity Measure}}
\label{sec:pairwise-similarity}
From the above definitions, we can observe that these two similarity measures only work for overlapping datasets. To address this, we introduce a more fine-grained measure, Hausdorff distance~\cite{Nutanong2011}, which considers the distance between each pair of points for two datasets.
 
\begin{definition1}{\textbf{(Hausdorff Distance (Haus)).}}
	\label{def:haus}
	{Given a query dataset $Q$ and a spatial dataset $D$, the Hausdorff distance from $Q$ to $D$ is the greatest of all the distances from any point $p \in Q$ to the closest point $p' \in D$, i.e.,}
	\begin{equation}
	{H}(Q, D) = \max_{p \in Q}{\min_{p'\in D}||p, p'||},
	\end{equation}
	\scalebox{0.95}{where $||p, p'||$ denotes the Euclidean distance between $p$ and $p'$}.
\end{definition1}

\subsection{Types of \underline{Dataset} Search}
\label{sec:query}
In what follows, we introduce two types of dataset search based on the query range and three aforementioned similarity measures, to help users find datasets of interest.

\begin{definition1}{\textbf{(Range-based Dataset Search  (RangeS)).}}
	\label{def:range}
	Given a rectangular range $R=[p_{min}, p_{max}]$,  {where $p_{min}$ and $p_{max}$ are the bottom left and upper right points to bound the range $R$, and a spatial data repository $\mathcal{D}$, a range-based dataset search returns all datasets $\mathcal{D}_q \subseteq \mathcal{D}$ such that: $\forall D \in \mathcal{D}_q$, the MBR of $D$ overlaps with $R$.} 
 
\end{definition1}


\begin{definition1}{\textbf{(Top-$k$ Exemplar Dataset Search (ExempS)).}}
	\label{def:topk}
	Given a query dataset $Q$ and a spatial data repository $\mathcal{D}$, a top-$k$ exemplar search returns $k$ datasets from $\mathcal{D}$ that are most similar to $Q$, i.e., it retrieves $\mathcal{D}_q \subseteq \mathcal{D}$ with $k$ datasets such that: $\forall D \in \mathcal{D}_q$ and $\forall D' \in \mathcal{D}\backslash \mathcal{D}_q$, we have $dist(Q,D) \leq dist(Q,D')$, where $dist$ is any of the distance measures defined in Section~\ref{sec:distanceMetrics}.
\end{definition1}

\subsection{Types of \underline{Data Point} Search}
\label{sec:enrich}    
After searching related datasets,  the goal of a more fine-grained query is to find the data point of interest.

\begin{definition1}{\textbf{(Range-based Point Search (RangeP)).}}
	\label{def:union}
	Given a query range ${R}$, {and a dataset $D \in \mathcal{D}_q$}, the range-based point search returns all the points of $D$ inside the range $R$.
 
 
\end{definition1}

\begin{definition1}{\textbf{(Nearest Neighbor Point Search (NNP)).}}
	\label{def:join}
	Given a query dataset $Q$, and a dataset $D \in \mathcal{D}_q$, the NNP search aims to find nearest points in $D$ for each point in $Q$.  
\end{definition1}


{
\subsection{Problem Formulation}
\label{sec:problem}
\begin{definition1}{\textbf{(Spatial Data Search Problem).}}
	\label{def:Spider}
We aim to propose a highly-efficient and scalable data search system that seamlessly integrates dataset and data point search with a unified index and several acceleration strategies, to work with different types of query types and distance measures.
\end{definition1}

\myparagraph{RoadMap}
The rest of this paper is organized as follows. We first introduce the overall system architecture in Section~\ref{sec:framework}. Then, we describe the unified index and how it is constructed in Section~\ref{sec:index}. Subsequently, we introduce a series of effective acceleration and pruning mechanisms in Section~\ref{sec:accelerating}. 
Finally, experimental results are provided in Section~\ref{sec:exp}.
}




\vspace{-0.2cm}
\section{Spatial Data Search System Framework}
\label{sec:framework}
\begin{sloppypar}

\begin{figure}
\setlength{\abovecaptionskip}{-0.10cm}
\setlength{\belowcaptionskip}{-0.10cm}
	\centering
	\includegraphics[width=7.7cm]{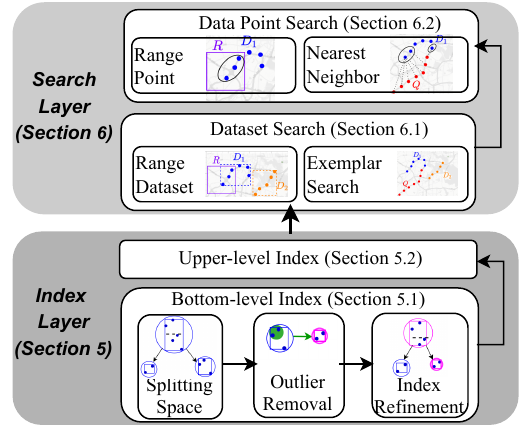}
\caption{Architecture of \spadas. }
	\label{fig:framework1}
\end{figure}

Given the problem described in Section~\ref{sec:defi}, we propose a multi-granularity \underline{spa}tial \underline{da}ta \underline{s}earch system called \spadas. The architecture of \spadas and its components are shown in Fig~\ref{fig:framework1}. It consists of two layers that we describe below.

\vspace{-.2cm}
\subsection{Index Layer}
Existing indexes mainly adopt a single-level index structure that only focuses on dataset-level or data point-level searches. To effectively support a rich set of queries presented in Section~\ref{sec:defi}, we design a novel two-level unified index structure in the index layer, including the upper-level index and bottom-level index, effectively organizing datasets and data points. 

\myparagraph{Bottom-level Index} 
The bottom-level index mainly focuses on the spatial distribution of data points within each dataset.
For the bottom index, we split the dataset node to create a \textit{dataset index} structure for fine-grained similarity measures and data point searches.
Considering that outliers can lead to imbalanced splits and degrade query performance, we first design \textit{parameter-free outlier removal} and \textit{index refinement} techniques to optimize the bottom-level index, which effectively removes outliers and improves the accuracy of similarity computations.


\myparagraph{Upper-level Index} 
The upper-level index is designed to partition and organize datasets at a higher level. Specifically, to avoid scanning all datasets and achieve a fast dataset search for a given query, we pre-organize the indexes of all datasets using a fast space-splitting indexing mechanism, such as the ball-tree \cite{Omohundro1989}. This allows us to construct an upper-level index, referred to as the \textit{data repository index}, which bounds all the dataset nodes. The index construction details are elaborated in Section~\ref{sec:index}.





\subsection{Search Layer}
\label{sec:searchlayer}
In the search layer, we support both coarse-grained dataset search and fine-grained data point search based on different queries and distance metrics. For dataset search, users have the option to specify a query region $R$ to identify all datasets overlapping with $R$. Alternatively, they can input a query dataset $Q$ to search for the top-$k$ datasets that have the minimum distance to $Q$.
After searching the datasets of interest, users can select one dataset for more fine-grained data point searches. These queries can help users find data points of interest to enrich the given query dataset. To accelerate all search operations, we design a set of pruning mechanisms based on the unified index at the search layer. In the following, we describe the basic intuition of pruning mechanisms.

\myparagraph{Pruning Mechanisms}
Based on building a two-level index, we design a set of unified pruning mechanisms to accelerate multiple types of search operations without constructing additional indexes. Specifically, when the query region $R$ or the query dataset $Q$ arrives, we traverse the index from the upper-level index to the bottom-level index in a depth-first manner. Then, we take the branch-and-bound strategy to batch prune out those nodes that cannot satisfy the query request, significantly speeding up the dataset search. The detailed pruning mechanisms are elaborated in Section~\ref{sec:accelerating}.

\end{sloppypar}





\vspace{-0.1cm}
\section{Optimization of Index Layer}
\label{sec:index}

\begin{figure}
	\centering
 \setlength{\abovecaptionskip}{0cm}
\setlength{\belowcaptionskip}{0cm}
\includegraphics[width=7.0cm]{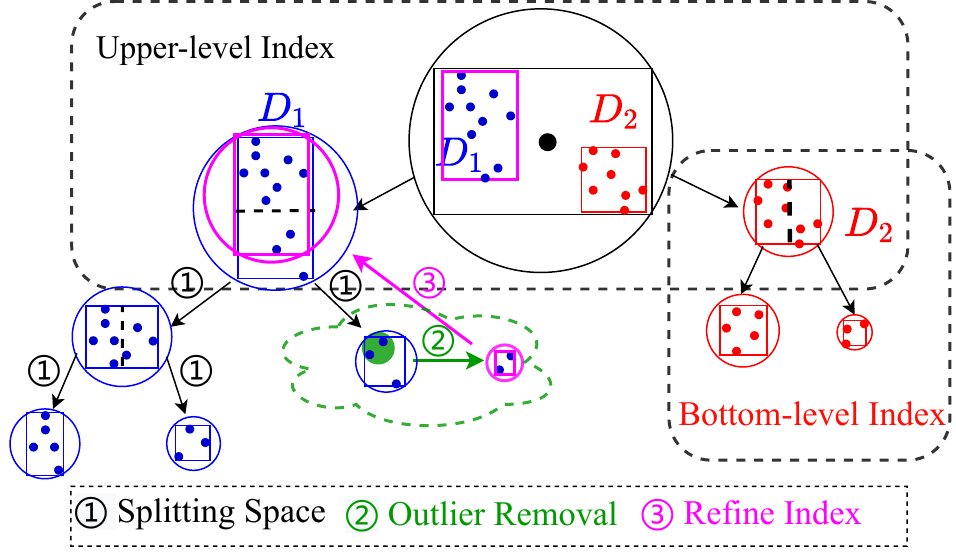}
	\caption{An overview of our unified index.}
	\label{fig:index}
\end{figure}





This section provides a detailed introduction to the index layer. Specifically, we design a unified index based on the whole spatial data repository (see Def.~\ref{def:datalake}), which includes a bottom-level index (see Section~\ref{sec:bottomindex}) and an upper-level index (see Section~\ref{sec:upperindex}). Figure~\ref{fig:index} presents an overview of our indexing framework. 
The index is built over a tree structure with nodes bounded by balls and boxes, and we store such bounding information to support multiple types of queries on spatial data. Algorithm~\ref{alg:indexing} shows the details about the implementation of index construction, which will be further explained in the following sections. 
In the subsequent subsections, we will introduce how the bottom-level (Section~\ref{sec:bottomindex}) and upper-level index (Section~\ref{sec:upperindex}) are constructed in detail.


%
\begin{algorithm}
\label{alg:constructIndex}
	\small
	\caption{\textsf{ConstuctIndex}($\mathcal{D}, f, d$)}
	\KwIn{$\mathcal{D}$: spatial data repository, $f$: leaf node capacity, $d$: dimension of spatial point}
	\KwOut{$N_{root}$: the root node of unified index}
	\label{alg:indexing}
	List of dataset root nodes $L\leftarrow \emptyset$\;
       Initial a sorted list $\phi\leftarrow \emptyset$; \Comment{\footnotesize{recording leaf nodes' radii}} \label{line:phi}
	
	\ForEach{$D_i\in \mathcal{D}$ }{
{ $N_{D_i} \leftarrow$\textsf{SplitSpace}($D_i$,$f$,$d$,$\phi$); 
\Comment{bottom-level index} \label{line:split}}
		
			
		$L.add(N_{D_i})$;
	}

	\textsf{OutlierRemoval}($L$, $\phi$); \Comment{parameter-free outlier removal }\label{line:remove}

	$N_{root} \leftarrow \textsf{SplitSpace}(L, f, d, \emptyset$);\Comment{upper-level index}\label{line:upperLevel}
	
	
	\KwRet ${N_{root}}$;
	
	\vspace{-0.5em}
	\noindent\rule{8cm}{0.4pt}

	{	
\SetKwFunction{FMain}{\textsf{SplitSpace}}
	\SetKwProg{Fn}{Function}{:}{}
	\label{alg:bottomlevelindex}
	\Fn{\FMain{$S, f, d, \phi$}}{
		\KwIn{{$S$: data point set or dataset node set}, $f$: capacity, $d$: dimension, $\phi$: sorted list of leaf node radius}
		\KwOut{$N_s$: dataset index node that bounds set $S$}
		\footnotesize{$max\_{width} \leftarrow -\infty$, $d_{split} \leftarrow 0$}\; 
  \label{line:splitBegin}
  \footnotesize{$leftChildList \leftarrow \emptyset, rightChildList \leftarrow \emptyset$\;}
		$N_s \leftarrow $ Generate a index node according to set $S$\;  \label{line:initial}
            \If{$S$ is a set of data points and $|S|\leq f$}{\label{line:leafBegin}
            $\phi.add(N_s.r)$\;
            \Return $N_s$\; \label{line:leafEnd}
            }\If{$S$ is a set of dataset nodes and $|S|\leq f$}{\label{line:leafBegin2}
            \Return $N_s$\;\label{line:leafend2}
            }
		\For{$i \leftarrow 1:d$}{
			\If{$N_s.b^{\uparrow}[i]-N_s.b^{\downarrow}[i]>$max\_{width}}{
				$d_{split} \leftarrow i$\;
    $max\_{width} \leftarrow N_s.b^{\uparrow}[i]-N_s.b^{\downarrow}[i]$;
			}
		}
		\ForEach{each object $s$ in $S$}{
  			\eIf{$s$ is a dataset node}{
				$temp \leftarrow s.o[d_{split}]$;
			}{
				$temp \leftarrow s[d_{split}]$; \Comment{$s$ is a data point}
			}
			\eIf{$temp>N_s.b^{\downarrow} [d_{split}]+\frac{max\_{width} }{2}$} 
			{
				$leftChildList.add(s)$\; \label{line:leftNode}
			}{
				$rightChildList.add(s)$\;  \label{line:rightNode}
			}
		}
		$N_s.N_l \leftarrow$ \textsf{SplitSpace}($leftChildList$, $f$, $d$, $\phi $)\; \label{line:splitLeftNode}
            $N_s.N_r \leftarrow$ \textsf{SplitSpace}($rightChildList$, $f$, $d$, $\phi $)\; \label{line:splitRightNode}
		\Return $N_s$; \label{line:splitEnd}
	}
}
	\vspace{-0.5em}
	\noindent\rule{8cm}{0.4pt}
	\SetKwFunction{FMain}{\textsf{OutlierRemoval}}
	\SetKwProg{Fn}{Function}{:}{}
	\label{line:outlierBegin}
	\Fn{\FMain{$L$, {$\phi$}}}{
		\KwIn{\footnotesize{$L$: list of dataset root nodes, {$\phi$: sorted list of leaf node radius}}}
	    
	  	
	  	
	  	$g{'} \leftarrow 0$, $i\leftarrow1$, $pos\leftarrow1$;
	  	
	  	\ForEach{$i$-th element in $\phi$}{ \label{line:phiCompute}
	  		$g_i \leftarrow$ Equation~\ref{equ:keed}, $i\leftarrow i+1$\; \label{line:phiComputeEnd}
	  		\If{$g_i>g{'}$\label{line:choosePhiBegin}}{$g{'}\leftarrow g_i$, $pos\leftarrow i-1$;}
  		}
  		$r{'}\leftarrow \phi[pos-1]$; \label{line:choosePhi}
	   
	    \ForEach{leaf root node $N_{D_i}$ in $L$}{ 
	        \textsf{RefineBottomUp}$(N_{D_i}, r')$; \label{line:outlierEnd}
	    }
	}

	\vspace{-0.5em}
	\noindent\rule{8cm}{0.4pt}
	\SetKwFunction{FMain}{\textsf{RefineBottomUp}}
	\SetKwProg{Fn}{Function}{:}{}
	\label{alg:refine}
	\Fn{\FMain{$N$, $r{'}$}}{
		\KwIn{$N$: dataset index node, $r{'}$: radius threshold}
		\eIf{$N$ is a leaf node}{\label{line:removeBegin} \label{line:refineBegin}
			\If{$N.r>r{'}$}{
			{	\ForEach{point $p$ in N}{
					\If{$||N.o, p||>r{'}$}{
						$N.L_p.remove(p)$;
					}}
				}
			}
			$N$.update$(o, r, b^{\uparrow}, b^{\downarrow})$; \label{line:removeEnd}
		}{
			 \textsf{RefineBottomUp}($N_r$, $r{'}$);\label{line:removeIteration1}
			 
			 \textsf{RefineBottomUp}($N_l$, $r{'}$);\label{line:removeIteration2}
			 
		}
	}
\end{algorithm}
\vspace{-0.1cm}
\subsection{Bottom Level: Index of Each Single Dataset}
\label{sec:bottomindex}
{
As shown in Figure~\ref{fig:index}, the bottom-level index is built for each dataset, primarily storing the distribution information of data points within each spatial dataset. Since outlier points in the dataset affected the accuracy of spatial dataset search, we also propose a parameter-free outlier removal technology in constructing the bottom-level index, which could effectively remove outliers in the dataset. In the upcoming sections, we will start by introducing the index nodes with multiple attributes to form the bottom-level index. Subsequently, we will introduce the process of bottom-level index building in detail. To construct the bottom-level index, we first transform each spatial dataset $D$ into a dataset root node $N_D$. 


\noindent\textbf{Dataset Root Node.}
    Given a spatial dataset $D$, a dataset root node $N_D$ corresponding to the dataset $D$ is denoted as a tuple $(id, z, L_p, o, r, N_l, N_r,   b^{\uparrow}, b^{\downarrow})$, where $id$ denotes the identifier of the dataset $D$, $z$ denotes the $z$-order signature (defined in Def.~\ref{def:zorder}) of $D$, $L_p$ is a list of nodes contained in dataset $D$, $o$ is the center point, $r$ is the radius, $N_l$ and $N_r$ are the left and right child nodes, and $b^\downarrow$ and $b^\uparrow$ are the bottom left and upper right corners of the dataset $D$'s MBR (defined in Def.~\ref{def:dataset}).

We compute the center point $o$ of $N_D$ by averaging the coordinates of all the points in $L_p$, and choose the farthest distance from $o$ to any point in $L_p$ as the radius $r$. $N_l$ and $N_r$ are the two child nodes generated during the construction of the index tree. In the bottom-level index, except the root node, internal nodes and leaf nodes are collectively referred to as the dataset index node, which is introduced as follows.

\noindent\textbf{Dataset Index Node.} A dataset index node {$N$ in the bottom-level index} is denoted as a tuple $(L_p, o, r, {N_l, N_r}, b^{\uparrow}, b^{\downarrow})$, where the meanings of the symbols in this tuple are consistent with those used in the dataset root node.


Algorithm~\ref{alg:indexing} presents the detailed procedure for constructing the bottom-level index, which is comprised of three steps: splitting space, parameter-free outlier removal, and bottom-up index refinement. Firstly, for each dataset $D \in \mathcal{D}$, we perform the function \texttt{SplitSpace} of Algorithm~\ref{alg:indexing} for each dataset (see Lines~\ref{line:splitBegin} to \ref{line:splitEnd}).

\myparagraph{Step1: Splitting Space} 
Given a spatial dataset $D$, we start by transforming the dataset into its corresponding root node (see Line~\ref{line:initial}). Then, we compute the width of the root node in each dimension and select the $d_{split}$-th dimension with the maximum width to split the root node into two child nodes (see Lines~\ref{line:leftNode} to \ref{line:rightNode}). Then, we continue to build the sub-tree for each of the two child nodes (see Lines~\ref{line:splitLeftNode} to \ref{line:splitRightNode}). On the contrary, if the size of data points in the dataset does not exceed the leaf node capacity $f$, the process of splitting space terminates (see Lines~\ref{line:leafBegin} to \ref{line:leafEnd}).

\myparagraph{Step2: Parameter-free Outlier Removal}
Most existing methods utilize the dataset index to remove outliers that have a limited number of neighbors within a threshold distance, and reconstruct the index \cite{Gupta2017,Tran2020a,Tran2016,Orair2010,Bhaduri2011}. Their common drawback is the high complexity and the requirement of specifying parameters such as the number of neighbors and threshold distance. Thus, we design a parameter-free mechanism to efficiently remove outliers in spatial datasets, utilizing the distribution of leaf node sizes in the dataset index (see Function \texttt{OutlierRemoval} of Algorithm~\ref{alg:indexing}). 

We observe in Figure~\ref{fig:radius} that there is only a small number of leaf nodes with a large radius in the index built for the two data repositories considered (see Section~\ref{sec:exp} for more details). We infer that the leaf nodes with large radius may contain outliers because outliers are normally far from other points in the datasets. Hence, a key question here is how to determine an appropriate radius to distinguish between nodes with or without outliers.

The main idea of our method is illustrated in Figure~\ref{fig:outlier}. Firstly, it computes the radii of all leaf nodes and sorts them in an array $\phi$. It then extends \textit{the Kneedle algorithm} \cite{Satopaa2011,kneed} to automatically select an appropriate threshold $r'$ based on the sorted array $\phi$. Specifically, it computes a gap ($g_i$) for a radius $\phi[i]$ as (see Lines~\ref{line:phiCompute} to \ref{line:phiComputeEnd}):
\begin{equation}\label{equ:keed}
\small
	g_i = \phi[0] - i \times \frac{\phi[0] - \phi[|\phi|-1]}{|\phi|}-\phi[i], i\in \{1, |\phi-1|\}. 
\end{equation}

Geometrically, the gap $g_i$ for a radius $\phi[i]$ computes the distance between the point ($i$, $\phi[i]$) and its projection on the segment connecting the first point ($0$, $\phi[0]$) and the last point ($|\phi|-1$, $\phi[|\phi|-1]$), as shown in Figure~\ref{fig:outlier}(b). The radius with the largest gap is selected as the threshold $r'$ (see Lines~\ref{line:choosePhiBegin} to \ref{line:choosePhi}), as it is the turning point where most leaf nodes start to have a similar radius.  Thus, leaf nodes with a radius greater than $r'$ are refined by removing potential outliers.

\begin{figure}
	\centering
 \setlength{\abovecaptionskip}{0 cm}
 \setlength{\belowcaptionskip}{0 cm}
	\includegraphics[width=0.23\textwidth]{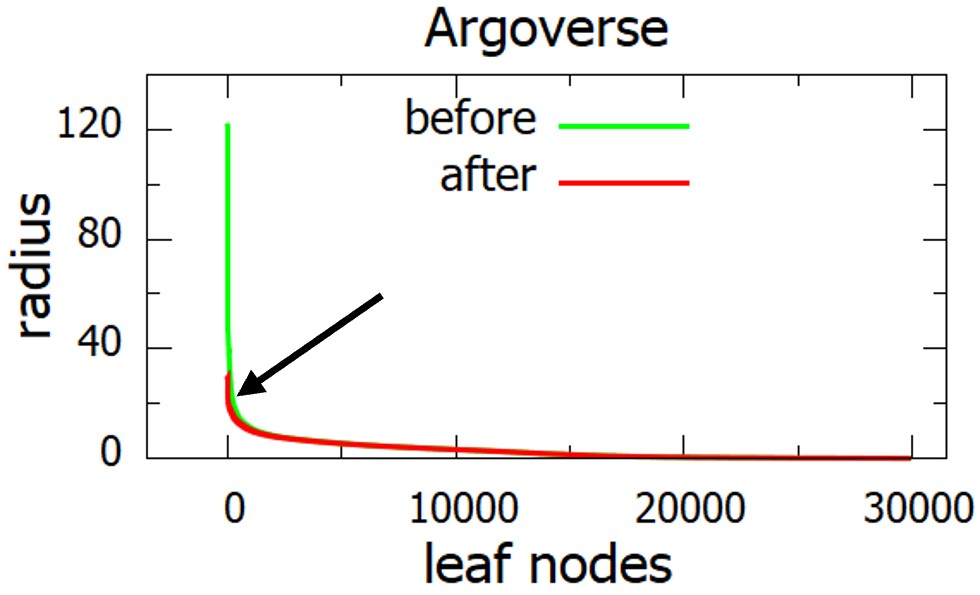}
	\includegraphics[width=0.23\textwidth]{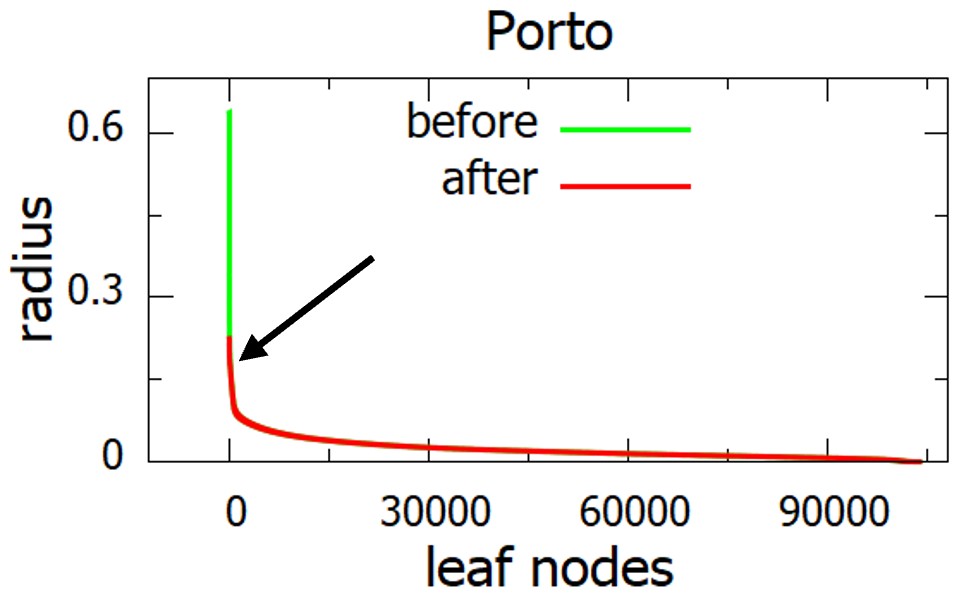}
	\caption{The distributions of radii of leaf nodes before and after outlier removal on two data repositories.}
	\vspace{-1.5em}
	\label{fig:radius}
\end{figure}

\begin{figure}
	\centering
	 \setlength{\abovecaptionskip}{-0.1 cm}
 \setlength{\belowcaptionskip}{-0.1 cm}
 \includegraphics[width=0.48\textwidth]{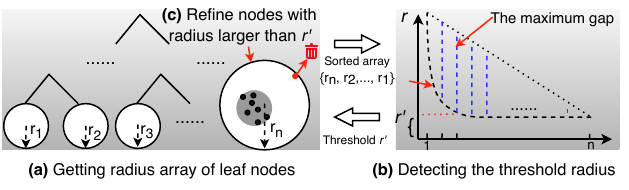}
	\caption{An illustration of the proposed parameter-free outlier removal method based on the Kneedle algorithm.}

	\label{fig:outlier}
\end{figure}

\myparagraph{Step3: Bottom-Up Index Refinement}
With the radius threshold $r'$, we traverse up from all leaf nodes to refine the dataset index until reaching the dataset root node in a dataset index. 
Specifically, for each leaf node, we remove all the points $p'$ that have a distance to the pivot point $p$ larger than $r'$ (see Lines~\ref{line:removeBegin} to \ref{line:removeEnd}), then refine the node and its upper-level nodes recursively in Function~\textsf{RefineBottomUp} (see Lines~\ref{line:removeIteration1} to \ref{line:removeIteration2}).
\subsection{Upper Level: Data Repository Index}
\label{sec:upperindex}
After constructing the bottom-level index, we further create an upper-level index in a similar top-bottom way to organize close nodes together. In the following, we will first introduce the index node in the upper-level index, followed by a detailed explanation of how the upper-level index is constructed.

\noindent\textbf{Data Repository Index Node.}
A data repository index node {$N$} is denoted as a tuple $(o, r, N_l, N_r, b^{\uparrow}, b^{\downarrow}, z, \delta)$, {where $\delta$ indicates whether $N$ is a leaf node, $z$ is the union of signatures of all its child nodes (i.e., $N.z=\bigcup_{N'\in N_l\cup N_r}N'.z$), and the meanings of other symbols in the tuple are the same as in the dataset root node.}

As shown in Figure~\ref{fig:index}, the upper-level index is constructed in a similar way to the bottom-level index (see Line~\ref{line:upperLevel}). Given a list of dataset root nodes $L$, we first initialize a dataset repository index node for the list $L$, and choose the widest $d_{split}$-th dimension to split the $L$ into two child lists based on their position in relation to the split line of parent node's MBR. Then, we create the dataset repository index node for each of the two child lists and continue to build the sub-tree, until the size of the child list satisfies the leaf node capacity $f$ (see Lines~\ref{line:leafBegin2} to \ref{line:leafend2}).

\myparagraph{Time Complexity}
Additionally, we provide a theoretical analysis of the time complexity for constructing the unified index, which is $O(m n\log\left(n\right))$, where $m$ and $n$ represent the number of data points in the dataset and the number of datasets in the data repository, respectively. The detailed analysis is provided in Appendix~\ref{appendix:indexComplexity}.

\myparagraph{Novelty of Unified Index}
In the process of index construction, previous indexes are designed for either spatial datasets or data points, which cannot support multiple types of searches. Thus, we design a two-level index structure, effectively organizing both datasets and data points, and can support spatial data search of different granularity at the same time. In addition, the existing indexes do not consider the influence of outliers in spatial datasets. Thus, we innovatively design outlier removal and index refinement technologies in index construction, to mitigate the influence of outliers on the accuracy of similarity measures between datasets.

\section{Optimization of Search Layer}
\label{sec:accelerating}

Based on the unified index constructed in the previous section, we propose effective pruning mechanisms to solve the data search problem defined in Section~\ref{sec:problem}. 

\subsection{\texorpdfstring{Accelerating Dataset Search}{Accelerating Dataset Search}}
\label{sec:accDataset}
In this subsection, we describe the dataset search algorithm applied to range-based dataset search and top-$k$ exemplar dataset search defined in Section~\ref{sec:query}. 
Algorithm~\ref{alg:top-k} shows the details of the implementation of the algorithm.



\subsubsection{\texorpdfstring{Accelerating Range-based Dataset Search}{Accelerating Range-based Dataset Search}}
\label{subsec:RangS}
Firstly, we introduce the accelerating strategy over range-based dataset search, as defined in Def.~\ref{def:range}, which only involves traversing the upper-level index to find datasets that overlap with the query region $R$. Specifically, we start from the root node of the upper-level index and traverse all its child nodes that intersect with $R$. Subsequently, we can directly prune the non-intersecting child nodes. For the intersecting child nodes, we iteratively traverse their child nodes if they are not leaf nodes; otherwise, we add their corresponding dataset IDs to the result list (see Lines~\ref{line:leafTraverse} to \ref{line:leafadd}). Finally, we return the result list that covers all datasets that overlap with $R$.


\begin{algorithm}[t]
\label{alg:datasetSearch}
\small
	\caption{\textsf{DatasetSearch}($Q$ or $R$, $\mathcal{D}$, $N$, $k$)}
	\KwIn{$Q$ or $R$: query dataset or range, $\mathcal{D}$: data repository, $N$: a node in the index,  $k$: number of results}
	\KwOut{$\mathcal{R}esult$: result list}
	\label{alg:top-k}
 $\mathcal{R}esult \leftarrow \emptyset$\;
	\eIf{$N$ is a dataset root node}{\label{line:leafTraverse}
		\eIf{\textbf{Range-based query}}{
			
			Add ${N.id}$ and update $\mathcal{R}esult$;\label{line:leafadd}
			
		}{
			\eIf{\textbf{Haus}}{
				$dis$ $\leftarrow$ \textsf{HausCompute}($Q$, $N$)\label{line:haus};
			}{
				\If{\textbf{IA} || \textbf{GBO}}{
					$dis$ $\leftarrow$ Compute $dis$ between $Q$ and $N$\label{line:iagbo};
				}
			}
			\If{$dis < \mathcal{R}esult[k]$}{
				
				Add ${N.id}$ and rank $\mathcal{R}esult$ increasingly by $dis$\label{line:iagbo2};
				
					
					
			}
		}	
	}
	{
		\ForEach{child node $N_c \in N.N_l \cup N.N_r$}{ \label{line:childTraverse}
			\eIf{\textbf{Range-based query}}{
				$\mathcal{R}esult$.add(\textsf{DatasetSearch}($R$, $\mathcal{D}$, $N_c$, $k$))\;
			}{
				$dis\leftarrow$ compute $dis$ between $Q$ and $N_c$\; \label{line:overlapDis}
				\If{$dis < \mathcal{R}esult$[k]}{
					$\mathcal{R}esult$.add(\textsf{DatasetSearch}($Q$, $\mathcal{D}$, $N_c$, $k$))\;
				}
			}
		}
	
	}
	
	\KwRet $\mathcal{R}esult$;
\end{algorithm}

\subsubsection{\texorpdfstring{Accelerating Top-$k$ Exemplar Dataset Search}{Accelerating Top-$k$ Exemplar Dataset Search}}
\label{sec:exemplarSearch}
To accelerate the top-$k$ exemplar dataset search defined in Def.~\ref{def:topk}, which supports two types of similarity queries: overlap-based queries (\IA  defined in Def.~\ref{def:ia} and \GBO defined in Def.~\ref{def:gbo}) and point-wise distance query (\Haus defined in Def.~\ref{def:haus}), we propose a series of acceleration strategies.
Compared to the other two measures, the Hausdorff distance has a quadratic computational complexity in terms of the number of points in the datasets. Thus, we propose a fast bound estimation technique to speed up the computation of the Hausdorff distance. In addition, to get the query results faster, users are often satisfied with approximate answers, especially if precision guarantees accompany these answers. Thus, we also propose a fast and efficient approximate Hausdorff distance computation with a theoretical error guarantee to balance efficiency and effectiveness.

\myparagraph{(1) Overlap-based Queries}
\label{overlapQuery}
In overlap-based queries, we have a ranking list that holds $k$ datasets with the maximum overlap.
Given a data repository node, we can prune it if (i) there is no overlap or (ii) the overlap-based distance (see Line~\ref{line:overlapDis}) is bigger than the $k$-th result. Taking the top-$k$ GBO search as an example, if a dataset repository node $N$ has fewer intersected grids with the query than the current $k$-th result, we can directly prune $N$ and all sub-trees of $N$, which greatly reduces the number of candidate datasets.




\myparagraph{(2) Point-wise Distance Query}
\label{PointWiseQuery}
As a point-wise distance measure, the computation time of Hausdorff increases dramatically with the number of points in the dataset since it has a quadratic time complexity.
To accelerate Hausdorff's computation, we propose a \textit{fast bound estimation} technique to efficiently estimate the lower and upper bounds of Hausdorff distance, as detailed below. 


\begin{figure}
\vspace{-0.3cm}
	\centering
 \setlength{\abovecaptionskip}{0 cm}
 \setlength{\belowcaptionskip}{0 cm}
	\includegraphics[width = 8.4cm]{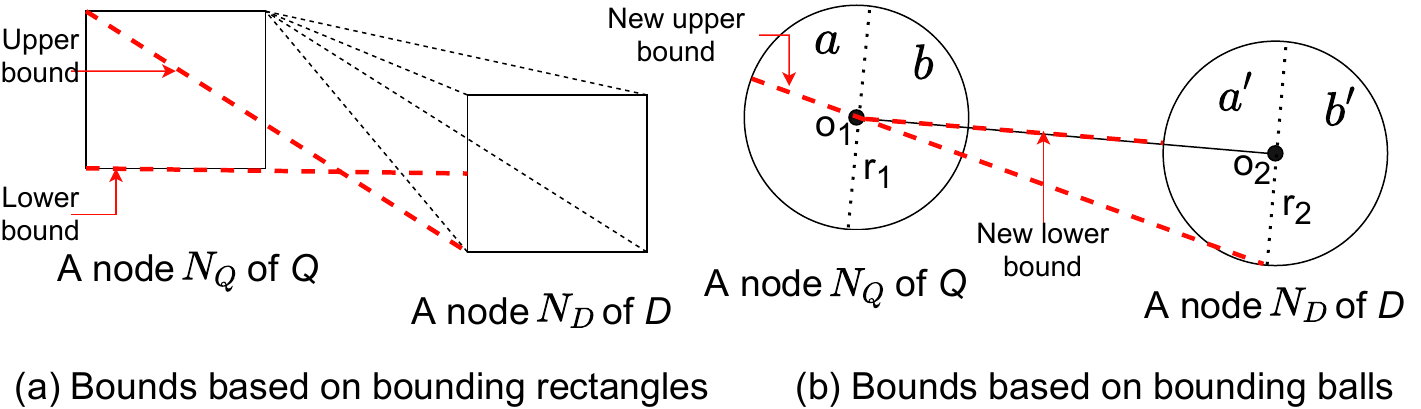}
	\caption{The distance bounds between two nodes using bounding boxes (left) and bounding balls (right).}
	\label{fig:ball-bound}
\end{figure}

\myparagraph{Fast Bound Estimation}
\label{sec:complexity}
We propose an efficient algorithm based on bounding balls of nodes to estimate the lower and upper bounds of the Hausdorff distance between nodes, which is not affected by the dimensions and only needs a single distance computation between center points. In contrast to the R-tree based incremental approach \cite{Nutanong2011}, which requires enumerating all pairs of corners of nodes' bounding boxes (as illustrated by the four black dotted lines in Figure~\ref{fig:ball-bound}(a)), our method avoids quartic Euclidean distance computations \cite{Nutanong2011} and only requires a single computation.

Next, we take Figure~\ref{fig:ball-bound}(b) as an example to elaborate on the lower and upper bounds of the Hausdorff we propose. When we construct an index node to bound all the points, for any diameter (see the two black dotted lines in Figure~\ref{fig:ball-bound}(b)) that divides the ball into two half-balls, each part should have at least one point. Otherwise, the ball should be refined to a ball with a smaller radius. We assume the two half balls are $a$ and $b$ of $Q$, $a'$ and $b'$ of $D$, and $b$ and $a'$ are two close half-balls. Since the Hausdorff distance is the greatest of all distances from any point $p\in Q$ to the closest point $p'\in D$, the closest point pair $(p, p')$ with the greatest distance in the Hausdorff distance computation from $Q$ to $D$ should be in $a$ and $a'$. We can easily get that $\forall p\in a, p' \in a'$, $||p, p'||\ge||o_1, o_2||-r_2$, and $||p, p'||\le\sqrt{||o_1, o_2||^2+r_2^2}+r_1$ according the triangle inequality. Still taking Figure~\ref{fig:ball-bound}(b) as an example, the lower bound should be the distance from $o_1$ to the closest point in the node $N_D$. The upper bound of the nearest neighbor will be the distance from the most left of $Q$ to the farthest in the left part of $D$. Hence, the Hausdorff distance $H(Q, D)$ between $Q$ and $D$ can be bounded in the range:
\begin{equation}\label{equ:haus}
 \small
    \left[\max(||o_1, o_2||-r_2, 0), \sqrt{||o_1, o_2||^2+r_2^2}+r_1\right].
\end{equation}

\begin{algorithm}[t]
\label{alg:hausCompute}
\small
	\caption{ \textsf{HausCompute}($Q$, $N_D$)}
	\KwIn{$Q$: query dataset, $N_D$: dataset root node}
	\KwOut{$UB$: Hausdorff distance between $Q$ and $D$}
	\label{alg:haus-join}
	$N_Q \leftarrow$ create a dataset root node of $Q$\;
	\If{\textbf{Haus}}{
		$PQ^{\uparrow} \leftarrow \textsf{CreateAscendingPQ}()$, $PQ^{\uparrow}.Insert(N_D, 0)$;
		
		$PQ^{\downarrow}\leftarrow \textsf{CreateDescendingPQ}()$, $PQ^{\downarrow}.Insert(N_Q, \infty, PQ^{\uparrow})$;
	}
	
	\While{$PQ^{\downarrow}.nonEmpty()$}
	{
	
	$[N_Q, UB, PQ^{\uparrow}] \leftarrow PQ^{\downarrow}.Dequeue()$;
	
	$[N_D, LB] \leftarrow PQ^{\uparrow}.Head()$,
	$terminate \leftarrow false$;
	
	\If{\textbf{Appro}\label{line:approx}}{
		\If{$N_D.r<\epsilon$ and $N_Q.r<\epsilon$ }{
                  $terminate \leftarrow true$\;
        }}

	\eIf{terminate}{
		\KwRet $UB$;
	}{
		\eIf{$N_D.r > N_Q.r$}{
			$\textsf{TraversalDIndex}(N_Q, PQ^{\uparrow}, PQ^{\downarrow})$;
		}{
			$\textsf{TraversalQIndex}(N_D, PQ^{\uparrow}, PQ^{\downarrow}, UB)$;
		}
	}
	}
	
	\vspace{-0.5em}
	\noindent\rule{8cm}{0.4pt}
	\SetKwFunction{FMain}{\textsf{TraversalQIndex}}
	\SetKwProg{Fn}{Function}{:}{}
	\label{alg:traversalx}
	\Fn{\FMain{$N_Q, PQ^{\uparrow}, PQ^{\downarrow}$}}{
		\KwIn{$N_Q$: query node, $PQ^{\uparrow}$: main queue, $PQ^{\downarrow}$: second queue}
		
		\ForEach{child $N_C$ in $N_Q$}
		{
			$CPQ^{\uparrow} \leftarrow CreateAscendingOrderPQ()$,
			$minUB\leftarrow \infty$;
			
			\ForEach{child node $N_C\in PQ^{\uparrow} $}
			{
				
				$CPQ^{\uparrow}.insert(N_C, \textsf{LowerBound}(N_C, N_D))$;
				
				$minUB \leftarrow \min(minUB, \textsf{UpperBound}(N_C, N_D))$;
			}
	
			$PQ^{\downarrow}.Insert(N_C, minUB, CPQ^{\uparrow})$;
		}
	}
	\vspace{-0.5em}
	\noindent\rule{8cm}{0.4pt}
	\SetKwFunction{FMain}{\textsf{TraversalDIndex}}
	\SetKwProg{Fn}{Function}{:}{}
	\label{alg:traversaly}
	\Fn{\FMain{$N_D, PQ^{\uparrow}, PQ^{\downarrow}, UB$}}{
		\KwIn{$N_D$: dataset node, $PQ^{\uparrow}$: main queue, $PQ^{\downarrow}$: second queue, $UB$: the upper bound}
		
		$[N_D, LB] \leftarrow PQ^{\uparrow}.Dequene()$, $minUB\leftarrow UB$;
		
		\ForEach{child $N_C \in N_D$}
		{	
			{
				$CPQ^{\uparrow}.insert(N_C, \textsf{LowerBound}(N_Q, N_C))$;
								
				$minUB\leftarrow \min(minUB, \textsf{UpperBound}(N_Q, N_C))$;
			}
		}
		$PQ^{\downarrow}.Insert(N_Q, minUB, PQ^{\uparrow})$;
	}
\end{algorithm}

\myparagraph{Exact Hausdorff}
\label{sec:exacthaus}
The above bounds can also be directly used in the fast pairwise Hausdorff distance computation.
The main idea is to rank candidates based on two priority queues.
Algorithm~\ref{alg:haus-join} shows how to prune distance computations based on bounds and achieve early termination without computing the full distance between all pairs of points from $Q$ and $D$. Similar to the incremental traversal (Please refer to Algorithm 4 in \cite{Nutanong2011})\footnote{Readers can refer to Algorithm 4 of \cite{Nutanong2011} for more details. Here, we will skip them as the estimation of the bounds is our main focus.}, since the goal of Hausdorff is to calculate the maximum nearest neighbor distance, we first create a priority queue in descending order, ranking the distance from the point to their nearest neighbors. Then to compute the minimum distances to the nearest neighbors for points in $Q$, we further create an ascending priority queue to traverse the dataset from close to far.
Subsequently, we traverse the index of $Q$ and $D$ from the root nodes, estimate the bounds between the polled pair of nodes, and conduct pruning. The whole algorithm terminates when the first pair of points are polled from the main priority queue. For convenience, we refer to this method as “\textbf{ExactAppro}” in Section~\ref{sec:exp}.

\myparagraph{Approximate Hausdorff}
\label{sec:approhaus}
Computing the exact Hausdorff distance can be time-consuming by going to the lowest point level of the dataset index. Instead, we can calculate an \textit{error-bounded Hausdorff distance} in a more efficient way based on the unified index.
In the exact distance computation, we terminate until a pair of points show up from the queue.
To avoid such intensive distance computations, we set a rule that once a pair of ``small'' nodes have a radius smaller than a given threshold $\epsilon$ (see Line~\ref{line:approx}), we compute their Euclidean distance between their center points.
Then we return the distance $||o_1, o_2||$ as the approximate distance, and do not enqueue their child nodes or points anymore. For convenience, we refer to this method as ``\textbf{ApproHaus}'' in Section~\ref{sec:exp}. Next, we prove that the \textbf{ApproHaus} method in Lemma~\ref{lem:error} achieves a guaranteed precision of $2\epsilon$, where $\epsilon$ is the error threshold defined by the user.
\begin{lemma1}
	\label{lem:error}
	Given a threshold $\epsilon$, the approximate algorithm guarantees an error of $2\epsilon$ for the Hausdorff distance.
\end{lemma1}

\begin{proof}
Based on the bound in Inequation~\ref{equ:haus}, we can get the upper bound distance and lower bound distance. 
When two nodes' radii are given, we can estimate the error bound based on the gap between their upper bounds and lower bounds:
\begin{small}
\begin{equation}
	\sqrt{||o_1, o_2||^2+r^2_2}+r_1 - ||o_1, o_2|| \le r_1+r_2< 2\epsilon,
\end{equation}
\end{small}
where we apply the triangle inequality:
\begin{small}
\begin{equation}
	\sqrt{||o_1, o_2||^2+r^2_2}- ||o_1, o_2|| \le r_2.
\end{equation}
\end{small}
For the lower bound, we can prove that:
\begin{small}
\begin{equation}
	\max(||o_1, o_2||-r_2, 0)-||o_1, o_2|| = -r_2\ge -\epsilon,
\end{equation}
\end{small}
which holds for both $||o_1, o_2||\ge r_2$ and $||o_1, o_2||\le r_2$.
Then, we proved that the error is bounded by $2\epsilon$. 
\end{proof}

\vspace{-0.5em}


To select a proper $\epsilon$, we can choose the cell size as the error threshold $\epsilon$ according to the MBR range as follows:
\begin{small}
\begin{equation}
	\label{equ:epsilon}
	\epsilon = \frac{b^\uparrow[0]-b^\downarrow[0]}{2^\theta}.
\end{equation}
\end{small}


\subsection{Accelerating Data Point Search}
\label{sec:accDataPoint}

\subsubsection{\texorpdfstring{Range-based Point Search}{Range-based Point Search}}
\label{subsec:RangeP}
Given a query range $R$ and a dataset $D$ from the query result $\mathcal{D}_q$, the range-based point search defined in Def.~\ref{def:union} aims to identify all points in $D$ that fall within the specified range. To expedite the range-based point search process, we employ a depth-first traversal approach starting from the root node $N_D$ of the dataset. During the traversal, we can prune the dataset index node $N$ if there is no overlap with the query rectangle $R$, i.e., $R \cap N.b =\emptyset$. Conversely, if there is an intersection, we proceed to evaluate the relationship between the node $N$ and $R$. When the query range $R$ entirely encompasses the node $N$, all data points beneath that node are added to the result list. For the intersecting node with $R$, we iteratively traverse their children until reaching leaf nodes. At each leaf node, we identify the specific points that fall within the range $R$ and append them to the result set.

\subsubsection{\texorpdfstring{Nearest Neighbor Point Search}{Nearest Neighbor Point Search}}
\label{subsec:NNP}
In nearest neighbor point search defined in  Def.~\ref{def:join}, for each point in the query set $Q$, we select a dataset $D$ from the query result $\mathcal{D}_q$ and find the nearest neighbor point within $D$.
A naive way to accelerate this is by using the nearest neighbor search with the dataset index, which is similar to the Hausdorff distance computation.
Since the Hausdorff queries have conducted the pruning of point pairs that are far, we can reuse the queues $PQ^{\uparrow}$ and $PQ^{\downarrow}$ during pairwise Hausdorff matching and further process based on the dataset index to accelerate the nearest neighbor point search.
With the cached priority queue in our Algorithm~\ref{alg:haus-join}, we keep en-queuing the node pair until the queue is empty, then every query point will find its nearest neighbor. 


\myparagraph{Novelty of Acceleration Strategies}
According to the links between these technological developments, we can see that the proposed acceleration strategies based on our two-level unified index can solve multiple types of search problems at the same time compared to previous methods, which can only solve a single search problem~\cite{LiFX12,Guttman1984,Bentley75}. In addition, we theoretically derive Hausdorff’s upper and lower bounds using the triangle inequality, and we explore the approximate Hausdorff value in depth with a guaranteed theoretical error.

\vspace{-0.2cm}
\section{Experiments}
\label{sec:exp}
We conduct experiments on six real-world dataset repositories to demonstrate the effectiveness and efficiency of \spadas. In Section~\ref{exp:setting}, we introduce the experimental setups. Then, we present the performance of our proposed unified index in Section~\ref{exp:index}. Next, we present the experimental results that vary with several key parameters to demonstrate the effectiveness and efficiency of the dataset search and data point search in Sections~\ref{exp:datasetDiscovery} and \ref{exp:dataPointDiscovery}, followed by our case study in Section~\ref{caseStudy}.


\begin{table}[t]
\centering
	\caption{Summary of six spatial data repositories.}
    \vspace{-0.3cm}
	\label{tab:datasets}
	\scalebox{0.9}{
\renewcommand\arraystretch{1.0}\addtolength{\tabcolsep}{-1pt}
 \begin{tabular}{cccc}
		\toprule
		\textbf{Data repository} & \textbf{\# of datasets} & \textbf{ \# of dimensions} &
        \textbf{Size (GB)}\\ \midrule
                    MultiOpen & 10,714 & 2  & 4.39 \\ 
		T-drive& 250,606&2 & 0.79 \\
		Argoverse & 280,000 & 3 & 17.87 \\
		ShapeNet& 32,135& 3 & 2.04 \\
		Chicago & 283,375& 11 & 1.99 \\
  		Porto& 1,400,000 &2  & 1.45 \\ 
		\bottomrule
	\end{tabular}}
\end{table}

\begin{table}
\centering
	\caption{Parameter settings.}
    \vspace{-0.3cm}
	\label{tab:parameters}
	\scalebox{0.9}{
\renewcommand\arraystretch{1.0}\addtolength{\tabcolsep}{3pt}
 \begin{tabular}{cc}
		\toprule
		\textbf{Parameter}	& \textbf{Setting} \\ \midrule
		$k$: number of results &\{\underline{10}, 20, 30, 40, 50\} \\ 
		$m$: number of datasets &\{$10^{-4}$, $10^{-3}$, $10^{-2}$, $10^{-1}$, \underline{1}\}$\times|\mathcal{D}|$ \\
		$\theta$: resolution &\{3, 4, \underline{5}, 6, 7\} \\
		$R$: query range &\{1, 2, \underline{3}, 4, 5\}$\times \epsilon$ \\
		$s$: \# of combined datasets & \{\underline{1}, 10, $10^2$, $10^3$, $10^4$\}\\
		$f$: leaf node capacity & \{\underline{10}, 20, 30, 40, 50\} \\
        $d$: dimension & \{\underline{2}, 4, 6, 8\} \\
		\bottomrule
	\end{tabular}}
\end{table}

\subsection{Settings}
\label{exp:setting}
\myparagraph{Datasets}
We conduct experiments on six representative spatial dataset repositories, each of which is downloaded from open-source spatial data portals. Table~\ref{tab:datasets} presents the scale and dimension of each data repository. Figure~\ref{fig:countHistogram} shows the distributions of the number of points in each data repository.
\begin{itemize}[leftmargin=*]
    \item \textbf{MultiOpen} dataset, a real-world spatial data repository comprising datasets collected from three different open-source platforms: Points of Interests (POIs) of Baidu Maps Open Platform~\cite{baidu}, BTAA Geoportal~\cite{btaa} and NYU Spatial Data Repository~\cite{nyu}. 
	\item \textbf{T-drive} dataset \cite{Yuan2010} contains a collection of real-world trajectory datasets generated by over 33,000 taxis in Beijing;
	\item \textbf{Argoverse} self-driving datasets \cite{ChangLS2019} which record five-second driving scenarios with thousands of points;
	\item \textbf{ShapeNet} \cite{Chang2015a}, large-scale datasets of shapes represented by 3D computer-aided design (CAD) models of objects;
	\item \textbf{Chicago} taxi trips \cite{chicago} dataset, which has 194,992,061 trips in total. As all the trips are quantized every 15 minutes, we split the whole dataset every 15 minutes to create a set of sub-datasets;
 	\item \textbf{Porto} dataset \cite{porto} transformed from 400+ taxis' trajectories in the city of Porto for one year.
\end{itemize}

\vspace{-0.2cm}
\begin{figure*}
 \setlength{\abovecaptionskip}{0cm}
\setlength{\belowcaptionskip}{-0.2cm}
	\centering
	\hspace{-1.5em}
         \includegraphics[width=2.6cm, height=2.1cm]{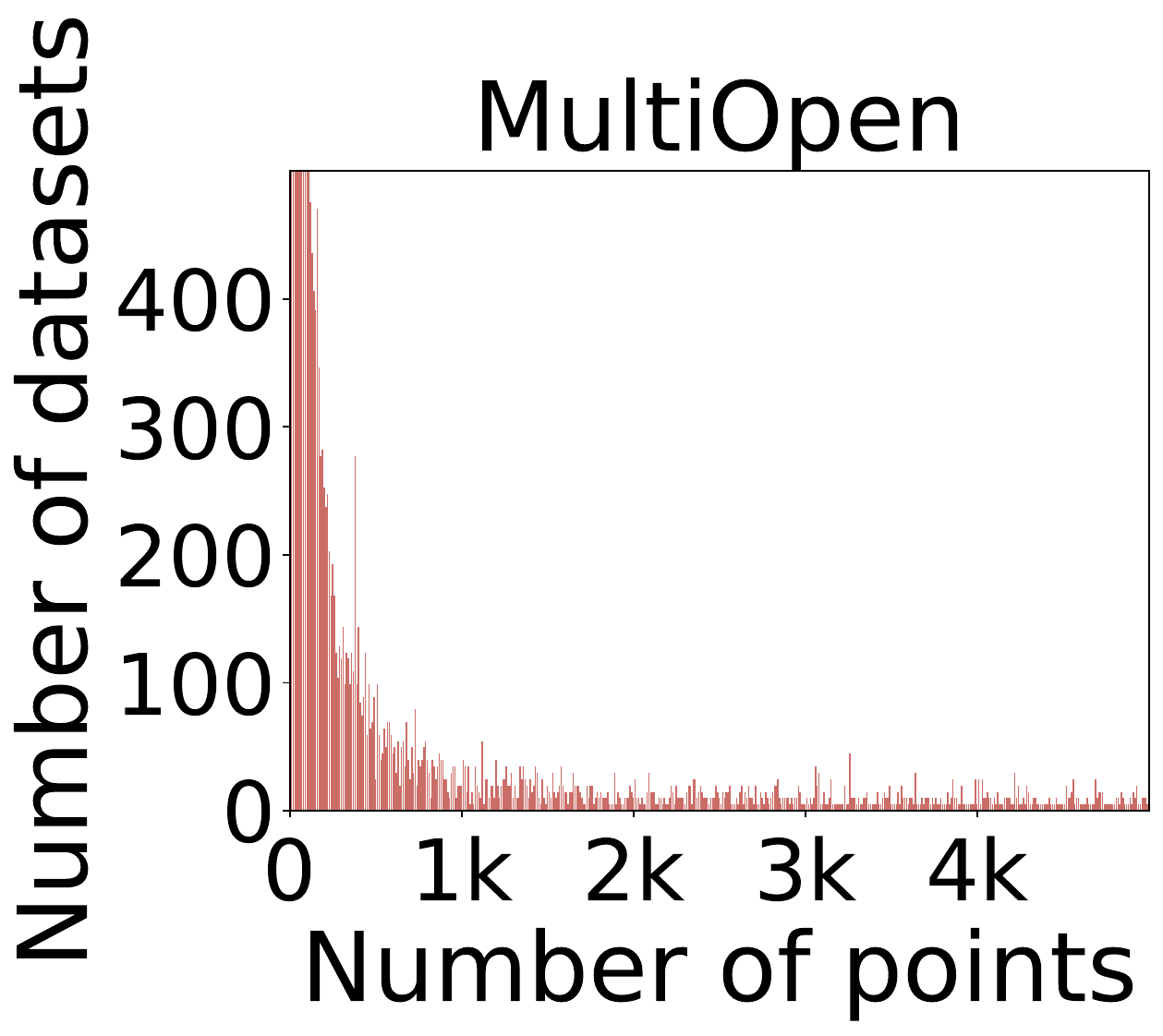}
	\includegraphics[width=2.6cm, height=2.1cm]{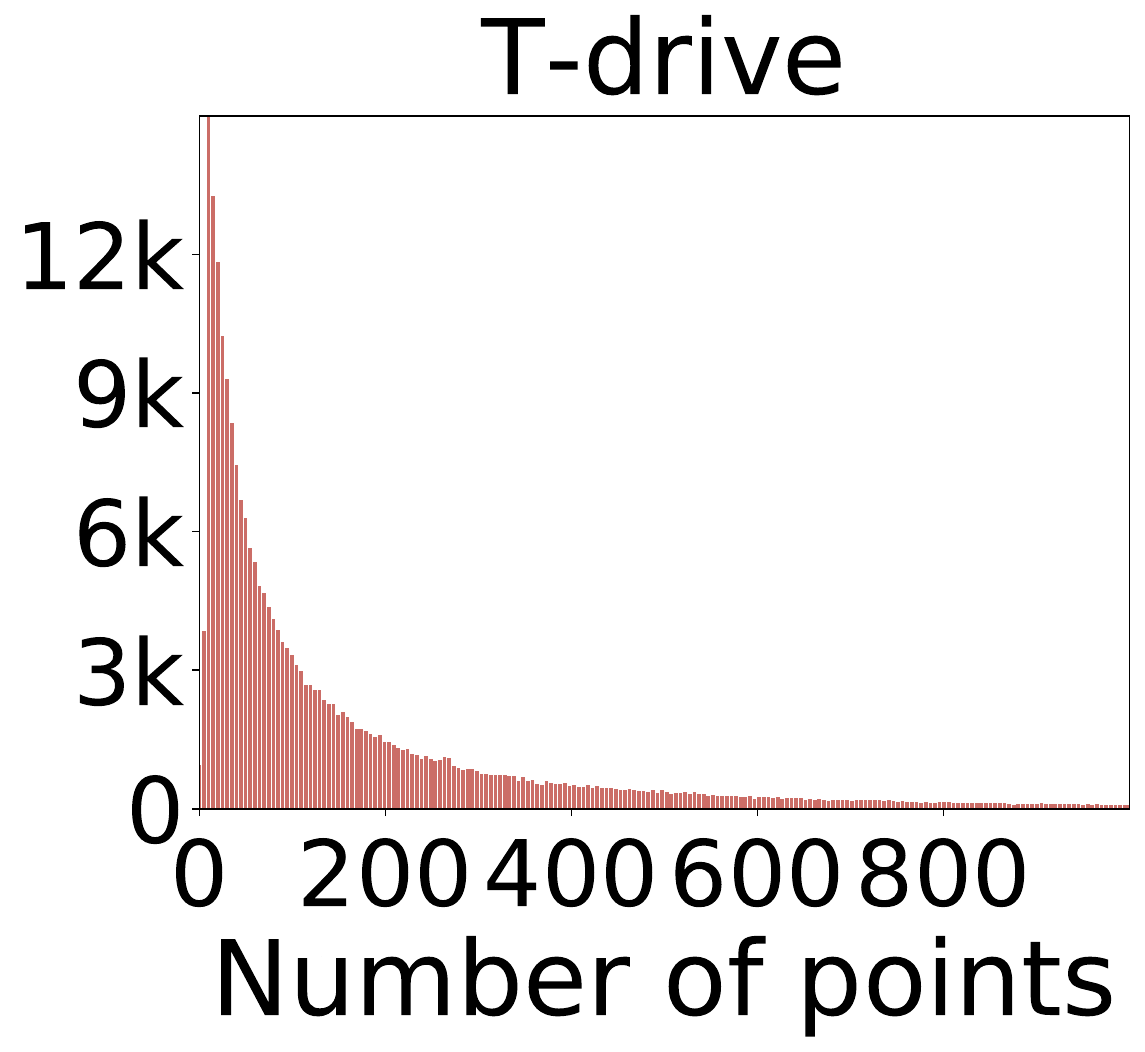}
	\includegraphics[width=2.6cm, height=2.1cm]{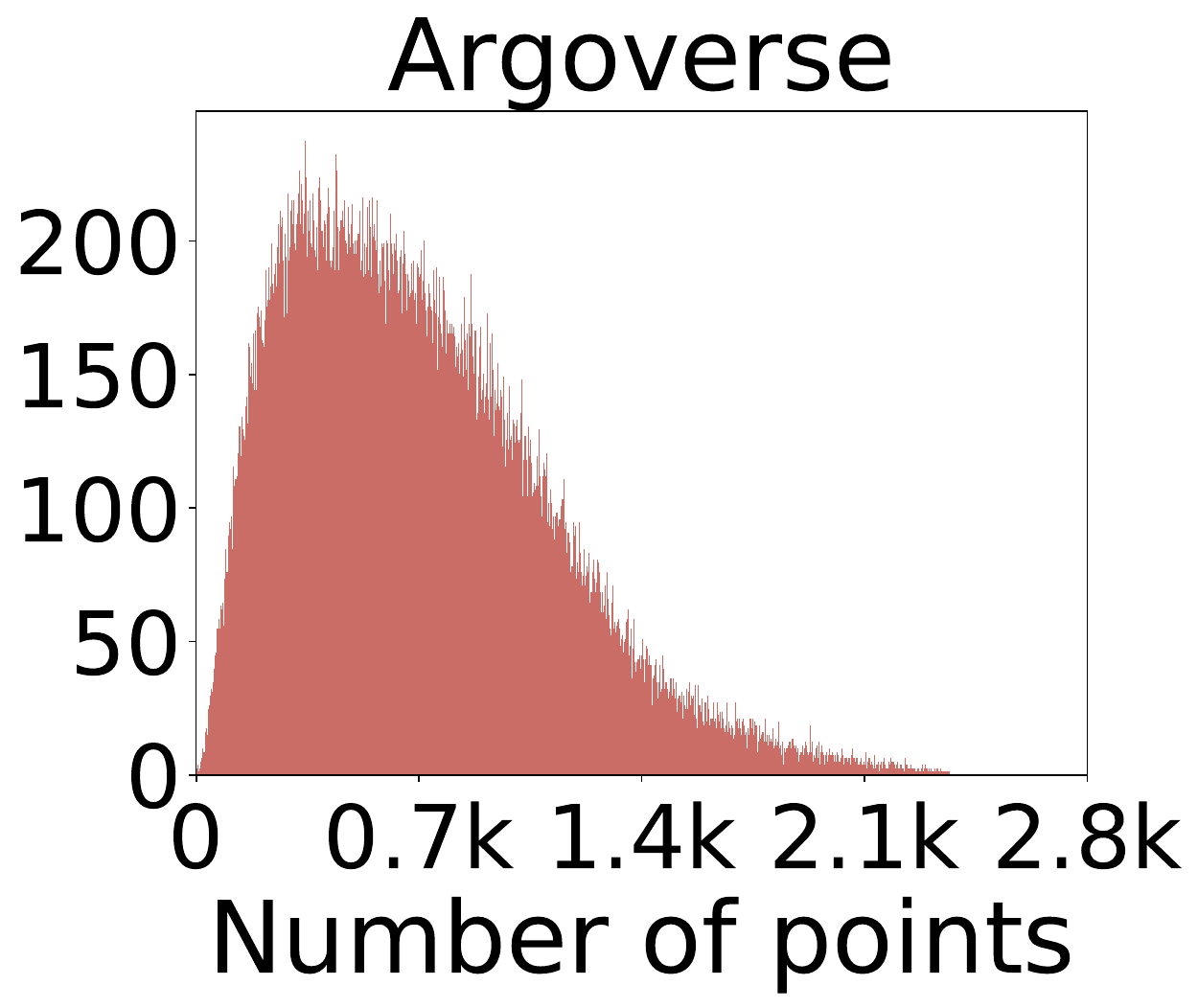}
	\includegraphics[width=2.6cm, height=2.1cm]{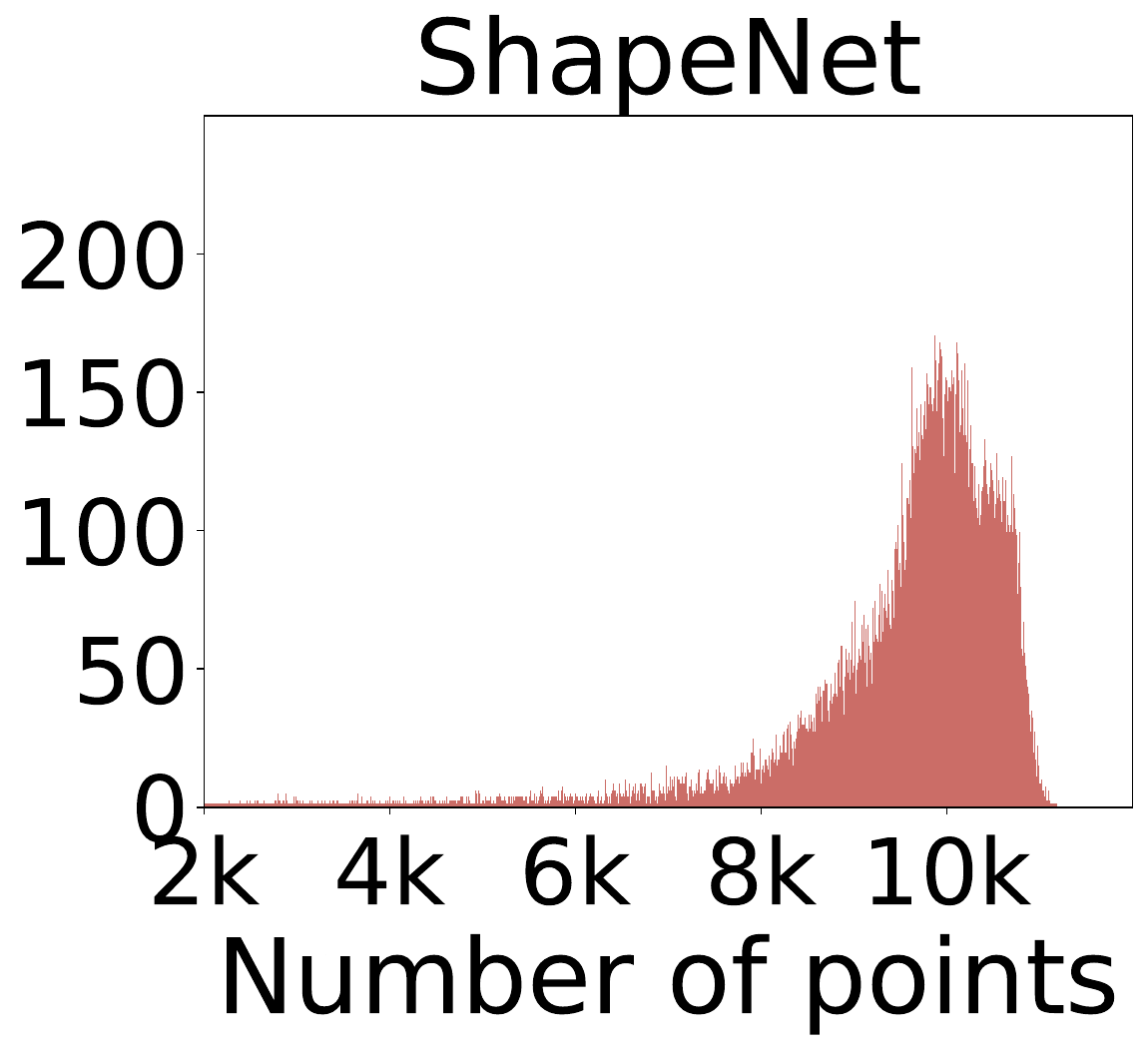}
	\includegraphics[width=2.6cm, height=2.1cm]{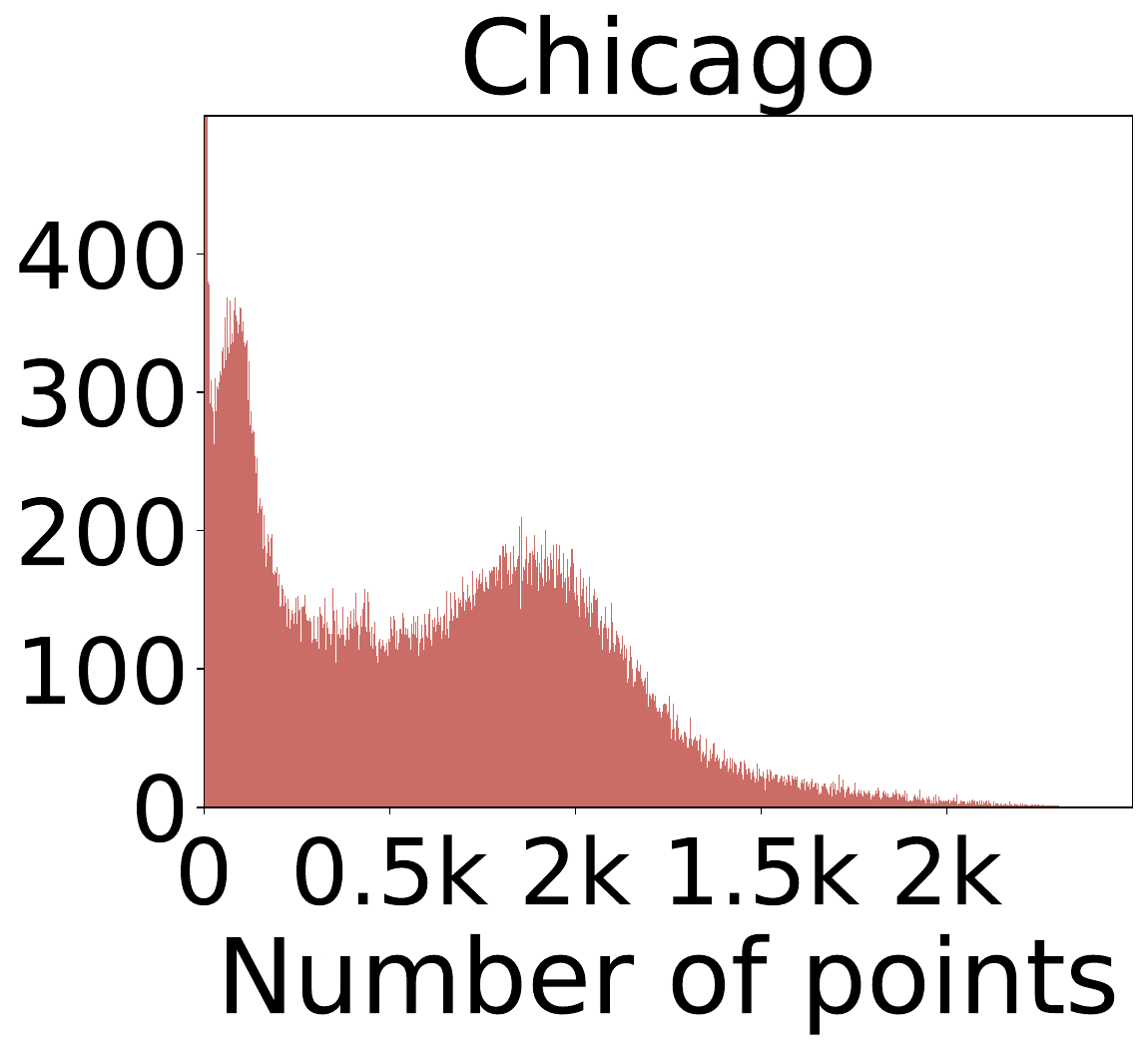}
         \includegraphics[width=2.6cm, height=2.1cm]{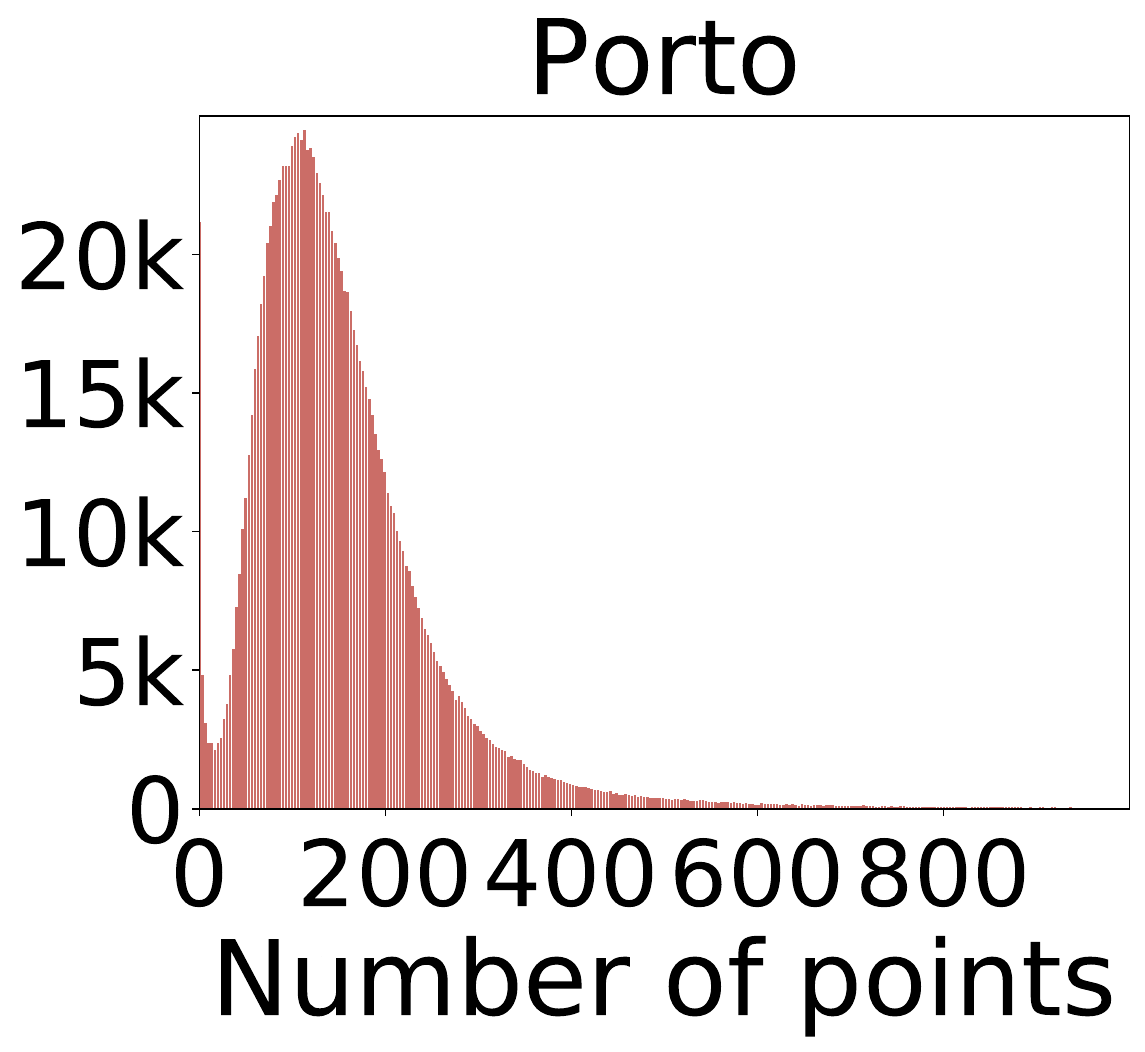}
\\
	\caption{Dataset scale distributions in six data repositories.}
	\vspace{-0.6em}
	\label{fig:countHistogram}
\end{figure*}


\myparagraph{Implementation}
We implement \spadas using Java 1.8 and run all our experiments on a 2x Intel Xeon Platinum 8268 24C 205W 2.9GHz Processor with 32G memory. Our code and guidelines for reproducibility can be accessed on the GitHub~\cite{spadas-code}. We also have an online demonstration system \cite{spade-system} on the Argoverse data repository, which supports the dataset and data point search proposed in this paper, and our proposed top-$k$ EMD search~\cite{YangWZ2022} is also available.

%

\myparagraph{Query Generation}
We randomly extract 100 datasets from each data repository as the query datasets to search the rest, and the final running time is logged on average from these 100 query datasets.
For the range-based query, we get the root node of each query dataset and infer the center point $o$, and we increase the number of cells to simulate the increasing range $R$ around $o$ to observe their effects on the performance. 


\myparagraph{Parameter Settings}
The settings of parameters are summarized in Table~\ref{tab:parameters} where the default settings are underlined. Below is the description of each parameter:
\begin{itemize}[leftmargin=*]
\item $k$: It specifies the number of results in the exemplar search.
\item $m$: It specifies the scale of the dataset in each data repository used to verify the scalability of our system.
\item $\theta$: It determines the size of cells in the grid.
\item $R$: It determines the range of query rectangular, where $\epsilon$ is the value obtained when the resolution $\theta$ is the default value $5$ according to the Equation~\ref{equ:epsilon} in Section~\ref{sec:exemplarSearch}.
\item $s$: It determines the scale of the query dataset in the nearest neighbor point (NNP) search (e.g., 10 means combining the points from the 10 query datasets into a larger dataset), which can be used to verify the scalability of the NNP.
\item $f$: It specifies the capacity of the leaf node, which is mainly used to determine when to terminate the index construction.
\end{itemize}

\vspace{-.1cm}
\myparagraph{Methods for Comparison}
 We compare our proposed \spadas system against the following baselines. The first four algorithms assess overlap-based dataset search, the next six evaluate pair-wise dataset search, and the last three test data point search. 

 \begin{itemize}[leftmargin=*]
 \item \IA (Def.~\ref{def:ia}): our top-$k$ exemplar dataset search based on the intersecting area in \spadas.
 \item \GBO (Def.~\ref{def:gbo}): our top-$k$ exemplar dataset search based on the grid overlap in \spadas.
  \item \RangeS (Def.~\ref{def:range}): our range-based dataset search in \spadas.
  \item \ScanGBO~\cite{Peng2016}: the baseline comparison method needs to scan the candidate datasets to compute the grid-based overlap for each query dataset (see Algorithm 2 in \cite{Peng2016}).
 \item \ExactHaus: our top-$k$ exemplar dataset search using exact Hausdorff distance with fast bound estimation.
 \item \ApproHaus: our top-$k$ exemplar dataset search using approximate Hausdorff distance with the error-bounded approximation.
  \item \ScanHaus~\cite{Adelfio2011}: It designs the lower and upper bounds of Hausdorff by using two point sets' MBR, and combines branch-and-bound to accelerate filtering out dissimilar datasets.
 \item \IncHaus~\cite{Nutanong2011}: It speeds up the Hausdorff distance computation by incrementally exploring the index of two point sets.
 \item \INNE \cite{bandaragoda2018isolation,zhao2019pyod}: the \textbf{i}solation-based anomaly detection using \textbf{n}earest-\textbf{n}eighbor \textbf{e}nsembles for outlier removal. 
   \item \Origin: the standard method for computing the exact Hausdorff distance without removing any outliers.
 \item \RangeP (Def.~\ref{def:union}): our range-based data point search in \spadas.
 \item \NNP (Def.~\ref{def:join}): our nearest neighbor point search in \spadas.
 \item \KNN~\cite{Taha2015}: a nearest neighbor search method using the early break
strategy.
 \end{itemize}


\vspace{-0.2cm}
\subsection{An Overview of spadas}
\label{exp:index}
We first show the running time of seven main steps in \spadas system, including the unified index construction (\textsf{Index}), dataset search (\RangeS, \IA, \GBO, \ExactHaus), and data point search (\RangeP, \NNP). As shown  in Figure~\ref{fig:overallacceleration}, we can observe that unified index construction {is the most time-consuming step} in most cases. However, the time taken to construct the index is still less than one minute. Moreover, the top-$k$ Hausdorff search is the slowest in dataset search compared with the \IA and \GBO due to its high complexity, but its running time is still tolerable. 

\myparagraph{Efficiency of the Unified Index}
We compare our unified index with the existing work that constructs an independent index using R-tree \cite{Nutanong2011} in \IncHaus. Figure~\ref{fig:indexConstructionN} presents the construction cost and the occupied space as the data repository scale $m$ increases. It is evident that our proposed unified index outperforms the state-of-the-art method in terms of both index construction time and space footprint. 
Specifically, with regard to index construction time, the unified index can achieve at most 36.8\% speedup over the index of \IncHaus. Additionally, in terms of index space, the unified index is reduced by up to 90\% over the \IncHaus.



\begin{figure}
	\centering
      \setlength{\abovecaptionskip}{0cm}
\setlength{\belowcaptionskip}{-0.1cm}
	\hspace{-1em}\includegraphics[width=7.5cm]{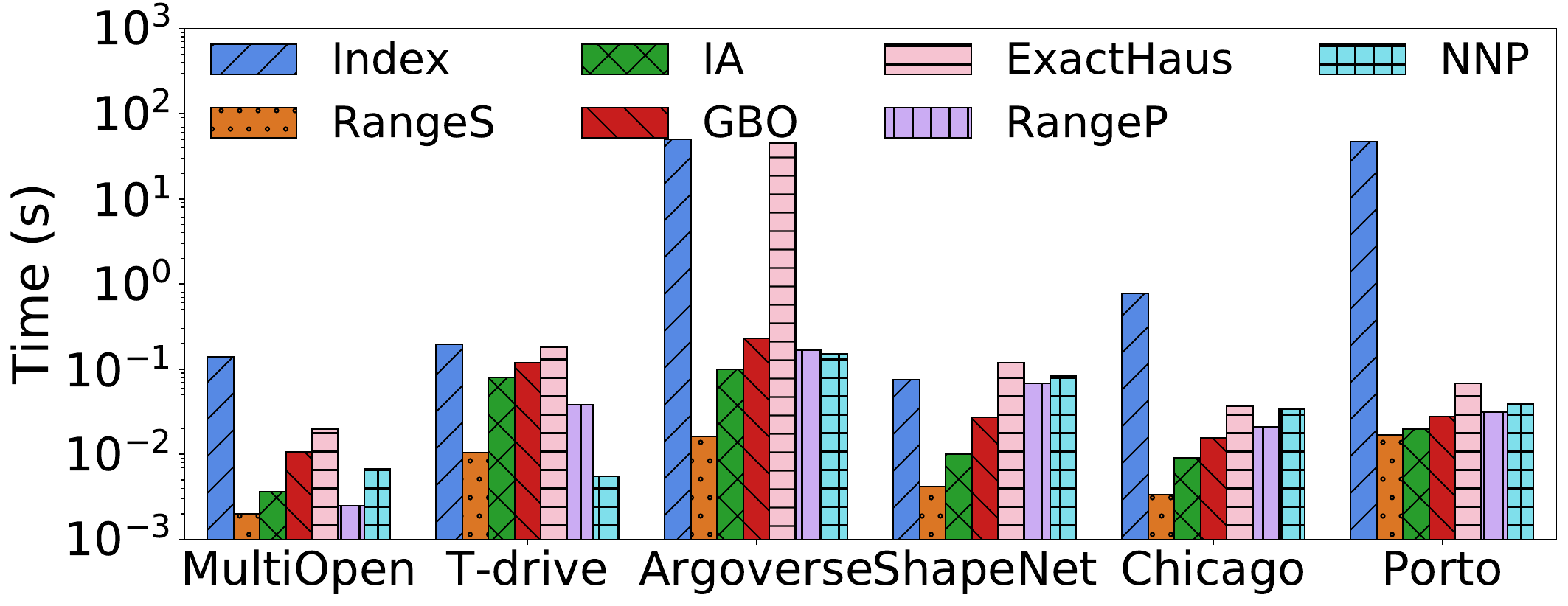}
	\caption{Efficiency in all stages on various data repositories.}
	\label{fig:overallacceleration}
\end{figure}

\vspace{-0.2cm}
\begin{figure*}
	\centering
  \setlength{\abovecaptionskip}{0cm}
\setlength{\belowcaptionskip}{-0.1cm}
	\hspace{-1em}\includegraphics[width=9cm]{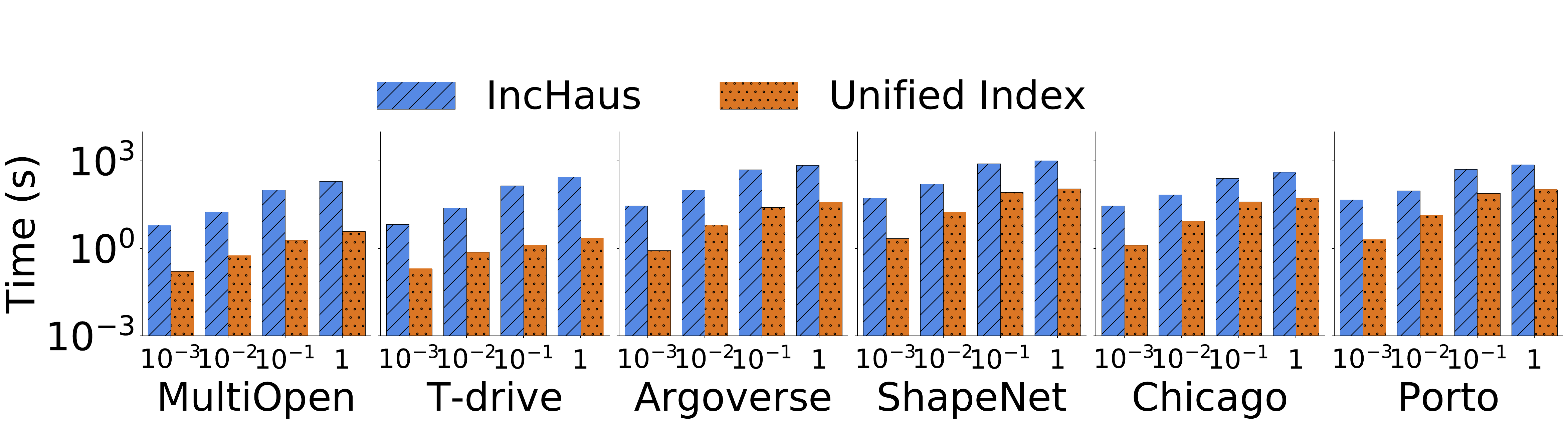}\includegraphics[width=9cm]{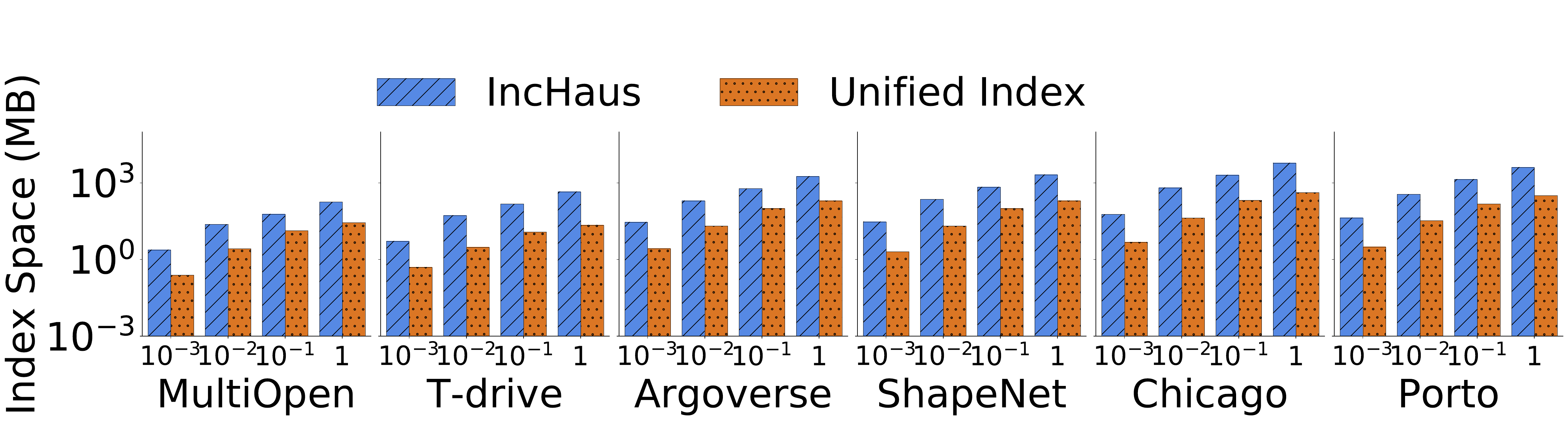}
	\caption{Index's construction time and occupied space with increasing data repository scale $m=\{10^{-3}, 10^{-2}, 10^{-1}$, 1\}$\times|\mathcal{D}|$.}
\vspace{-1.0em}
	\label{fig:indexConstructionN}
\end{figure*}

\begin{figure*}[t]
 \setlength{\abovecaptionskip}{0cm}
\setlength{\belowcaptionskip}{-0.1cm}
 \begin{minipage}[b]{.49\textwidth}
  \centering
  \hspace{-1.7em}
 \includegraphics[width = 9cm]{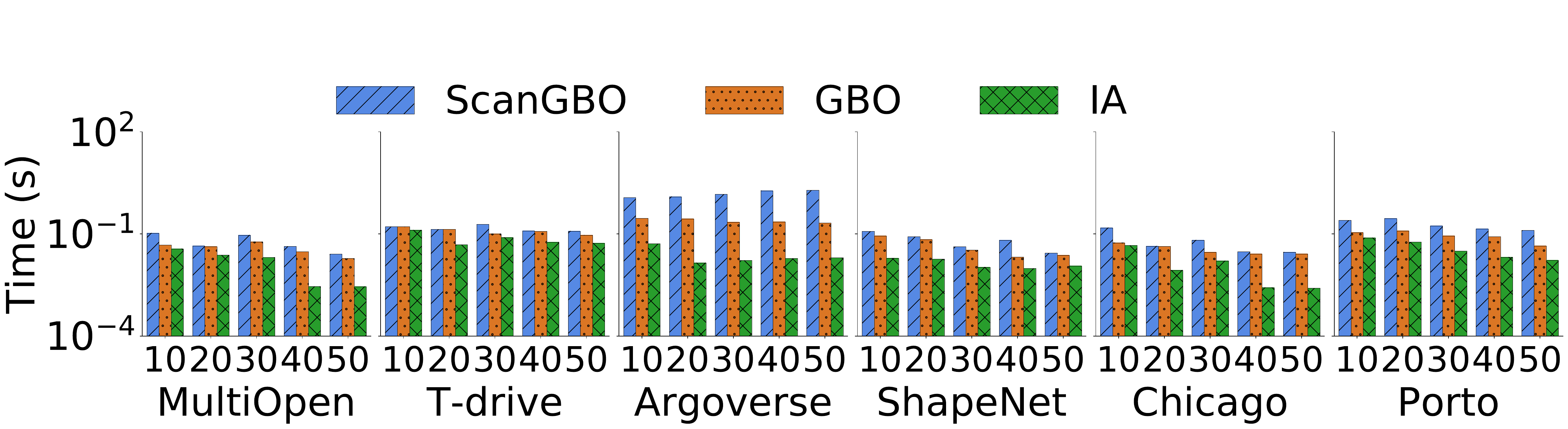}
  \caption{Top-$k$ overlap-based queries as $k$ increases.}
\label{fig:topkOverlapquery}
 \end{minipage}
 \begin{minipage}[b]{.49\textwidth}
  \centering
  \includegraphics[width = 9cm]{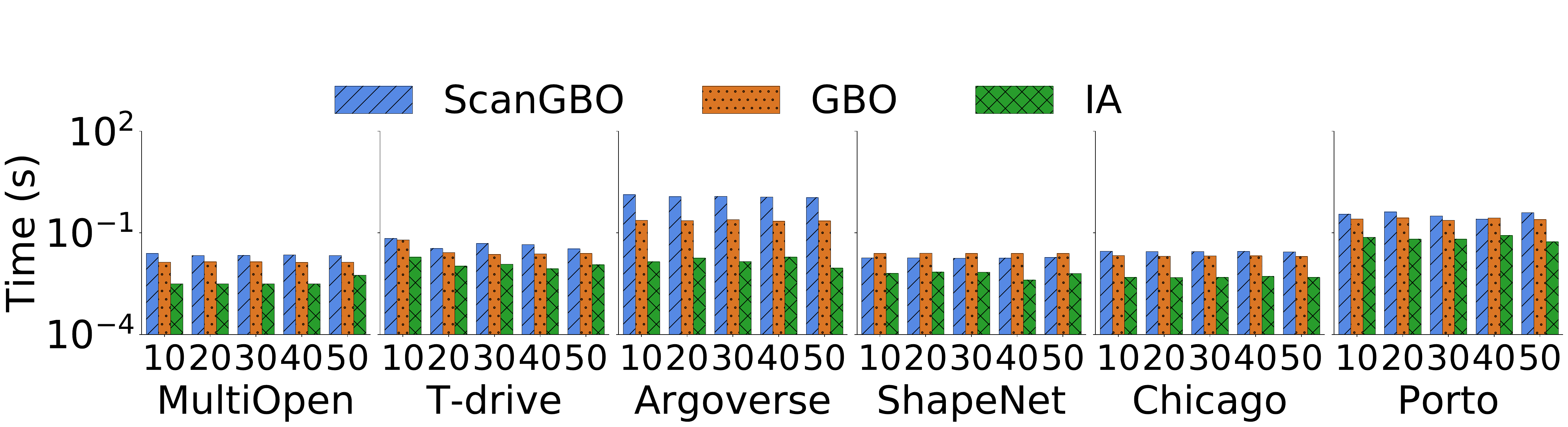}
\caption{Top-$k$ overlap-based queries as $f$ increases.}
\label{fig:topkOverlapF}
 \end{minipage}
 \vspace{-1.0em}
\end{figure*}





\begin{figure*}[t]
 \setlength{\abovecaptionskip}{0cm}
\setlength{\belowcaptionskip}{0cm}
 \begin{minipage}[b]{.49\textwidth}
  \centering
  \hspace{-1.7em}
 \includegraphics[width = 9cm]{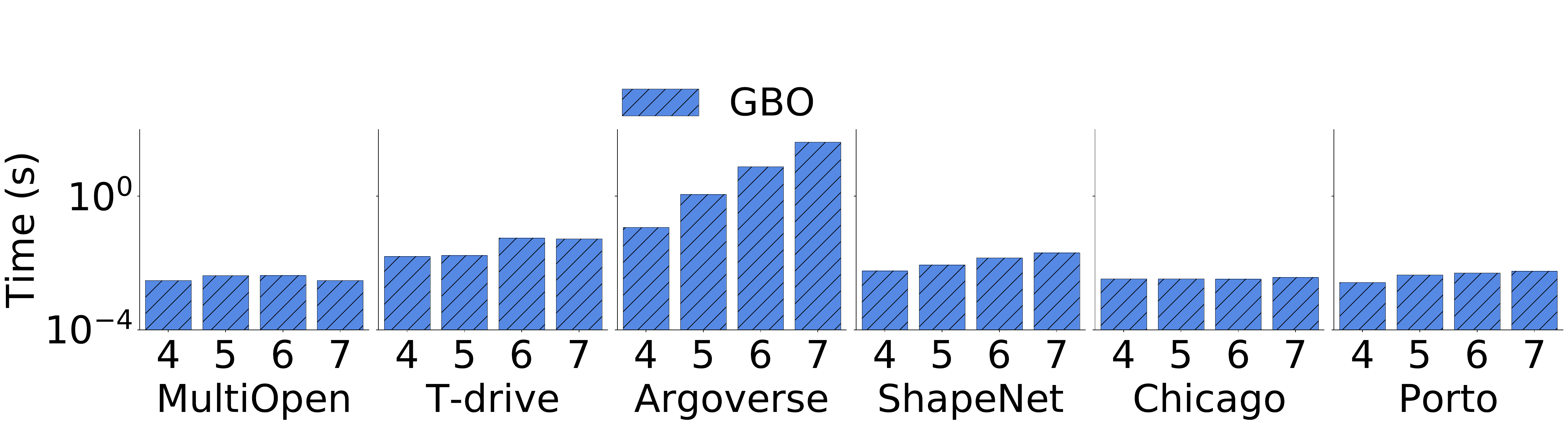}
		\caption{Grid resolution $\theta$'s effects on top-$k$ GBO query.}
\label{fig:topkGBOquery}
 \end{minipage}
 \begin{minipage}[b]{.49\textwidth}
  \centering
 \includegraphics[width = 9cm]{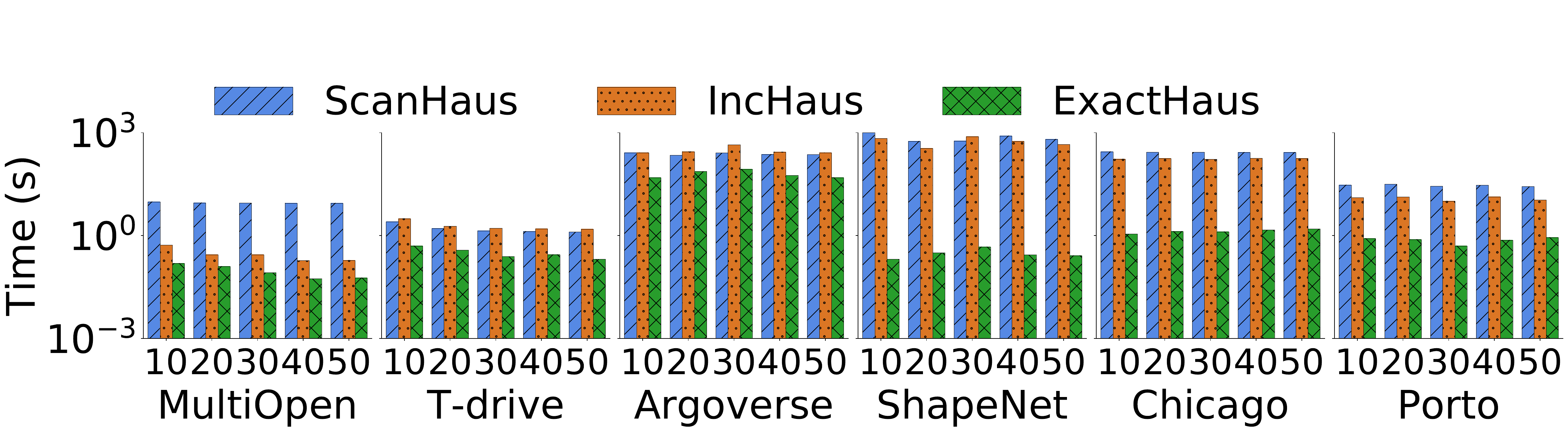}
\caption{Top-$k$ exact Hausdorff search with increasing $k$.}
\label{fig:topkHausA}
 \end{minipage}
 \vspace{-1.6em}
\end{figure*}

\begin{figure}
		\centering
     \setlength{\abovecaptionskip}{0 cm}
\setlength{\belowcaptionskip}{0 cm}
  \hspace{-1em}
\includegraphics[width=9cm]{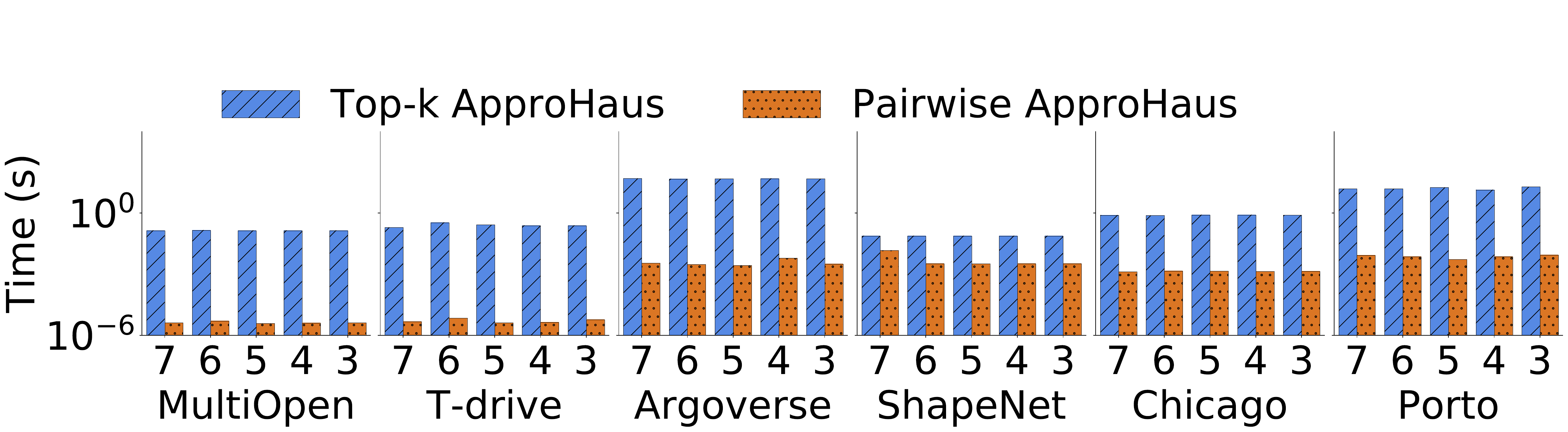} 
		\caption{Top-k approximate Hausdorff search with increasing threshold $\epsilon$ (i.e., decreasing resolution $\theta$ in x-axis).}
		\label{fig:topkHausB}
\end{figure}

\subsection{\textbf{Efficiency of Dataset Search}}
\label{exp:datasetDiscovery}
As the range-based dataset search is very fast (see Figure~\ref{fig:overallacceleration}), we mainly test the performance of the top-$k$ exemplar search.

\subsubsection{\textbf{Top-$k$ Overlap-based Query}}
\label{sec:topKoverlap}
We vary several key parameters to observe the search performance of overlap-based search.

\myparagraph{Effect of Number of Results $k$}
 As shown in Figure~\ref{fig:topkOverlapquery}, we increase $k$ to observe the search performance of coarse metrics based on overlaps. We can observe that the \IA is the fastest compared to the other two measures, because it only needs to compute the intersection area, while \GBO needs to conduct an intersection on integer sets. In addition, \GBO consistently outperforms the \ScanGBO, which utilizes a sequential scanning approach. Specifically, the \GBO can achieve up to at most 53\% times speedups over the \ScanGBO, which demonstrates the effectiveness of our method.


\myparagraph{Effect of Leaf Node Capacity $f$} We show the results of overlap-based search with the increase of leaf node capacity $f$. From Figure~\ref{fig:topkOverlapF}, we can observe that the search time of the three methods does not change significantly with the increase of $f$, indicating that the leaf node capacity during index construction has little influence on the search efficiency. In addition, we can see that \GBO and \IA consistently outperform the \ScanGBO, indicating that the pruning mechanism based on the unified index is very efficient.

\myparagraph{Effect of Grid Resolution $\theta$}
 Figure~\ref{fig:topkGBOquery} shows the performance of \GBO as the resolution $\theta$ increases. Since increasing $\theta$ leads to a higher number of cells spanning the entire space, the cell size is reduced. Thus, this enlarges the $z$-order signature in each index node, resulting in a degradation of efficiency. {Specifically, due to the relatively large size and tight distribution of the Argoverse dataset, increasing $\theta$ leads to more pronounced increases in z-order signature, resulting in a greater search time.}



\vspace{-.1cm}
\subsubsection{\textbf{Top-$k$ Point-wise Query}}
\label{sec:topKpointwise}
We vary several key parameters to test the performance of the exact and approximate Hausdorff.

\myparagraph{Effect of Top-$k$ Search} 1) \textbf{Exact Hausdorff}. We compare the top-$k$ Hausdorff search performance of our \ExactHaus with \ScanHaus and \IncHaus as $k$ increases. As shown in Figure~\ref{fig:topkHausA},
we can observe that our \ExactHaus based on the unified pruning mechanism drastically improves the performance of top-$k$ Hausdorff search, achieving up to at most three orders of magnitude speed improvements against the \ScanHaus and \IncHaus. This proves that our fast bound estimation technology is effective. 




\noindent2) \textbf{Approximate Hausdorff}. The approximate Hausdorff is further evaluated by decreasing the error threshold $\epsilon$. Since the error threshold increases as the resolution decreases, we present the experimental results of our proposed \ApproHaus in pairwise Hausdorff computation (i.e., computing the Hausdorff distance between two spatial datasets) and top-$k$ Hausdorff search, as the resolution decreases, in Figure~\ref{fig:topkHausB}.
We observe that the runtime of pairwise \ApproHaus computation degrades slightly with the increase of the error threshold. This is because in order to achieve a smaller error, the algorithm needs to scan more child nodes in the dataset index for approximate Hausdorff computation and terminate later.

{Furthermore, as illustrated in Figure~\ref{fig:resultHausExactAppro}, we compare the effectiveness and efficiency of top-$k$ \GBO and \Haus search in two modes: \ApproHaus and \ExactHaus. We use the top-$k$ search results of \ExactHaus as the ground truth (with a constant accuracy of 1 in Figure~\ref{fig:resultHausExactAppro}), and the accuracy of \ApproHaus is defined as the number of identical datasets between its top-$k$ results and the top-$k$ results of \ExactHaus divided by $k$.
The experimental results show that although \GBO is faster than the two Hausdorff searches, the accuracy is much lower (less than 80\%). In contrast, our \ApproHaus algorithm can achieve up to at most 81.6\% times speedups over \ExactHaus on various data repositories and maintain high accuracy (up to 90\%). This proves that our proposed error-bounded approximation technology is effective.}


\begin{figure*}[t]
 \setlength{\abovecaptionskip}{-0.1cm}
\setlength{\belowcaptionskip}{-0.1cm}
 \begin{minipage}[b]{.49\textwidth}
  \centering
\subfigure[$\!$Accuracy of Top-$k$ $\!$Hausdorff Search]{\includegraphics[width=4.2cm]{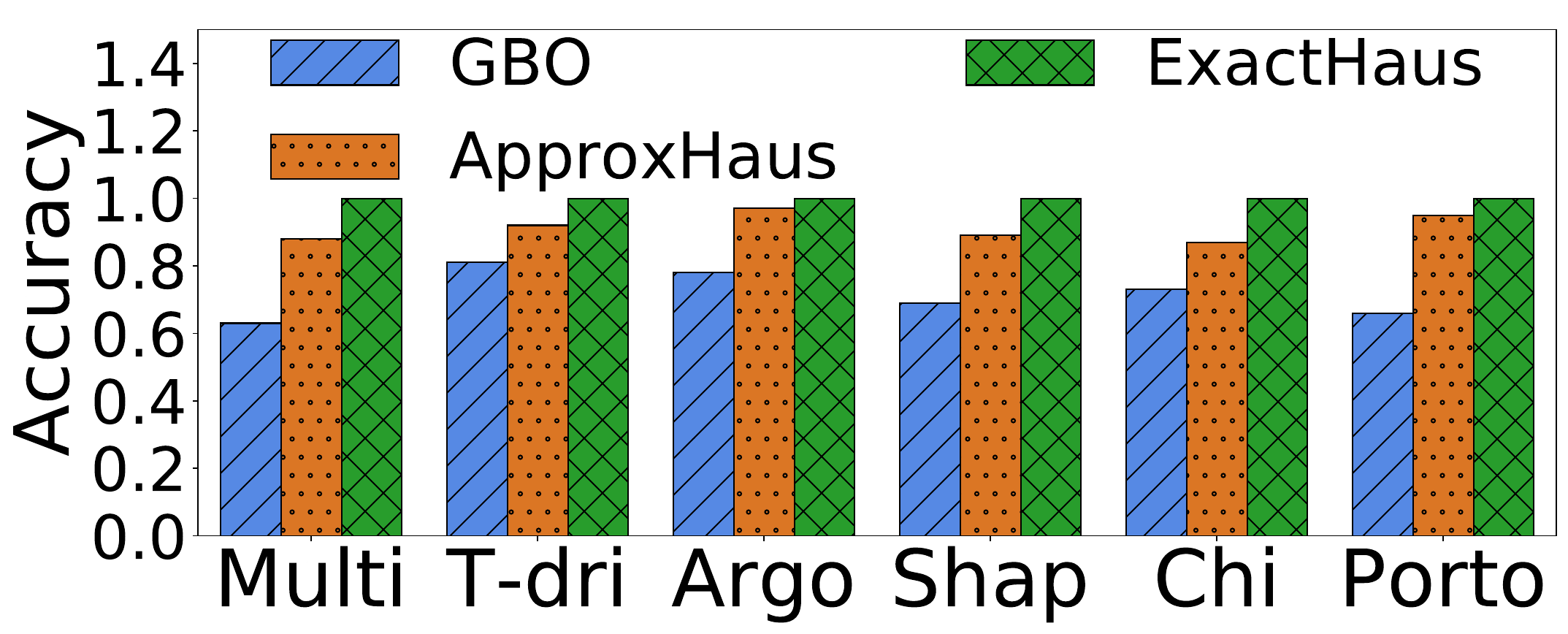}\vspace{-2.3cm}}
 \hspace{0cm}
\subfigure[Runtime of Top-$k$Hausdorff Search]{ \includegraphics[width=4.2cm]{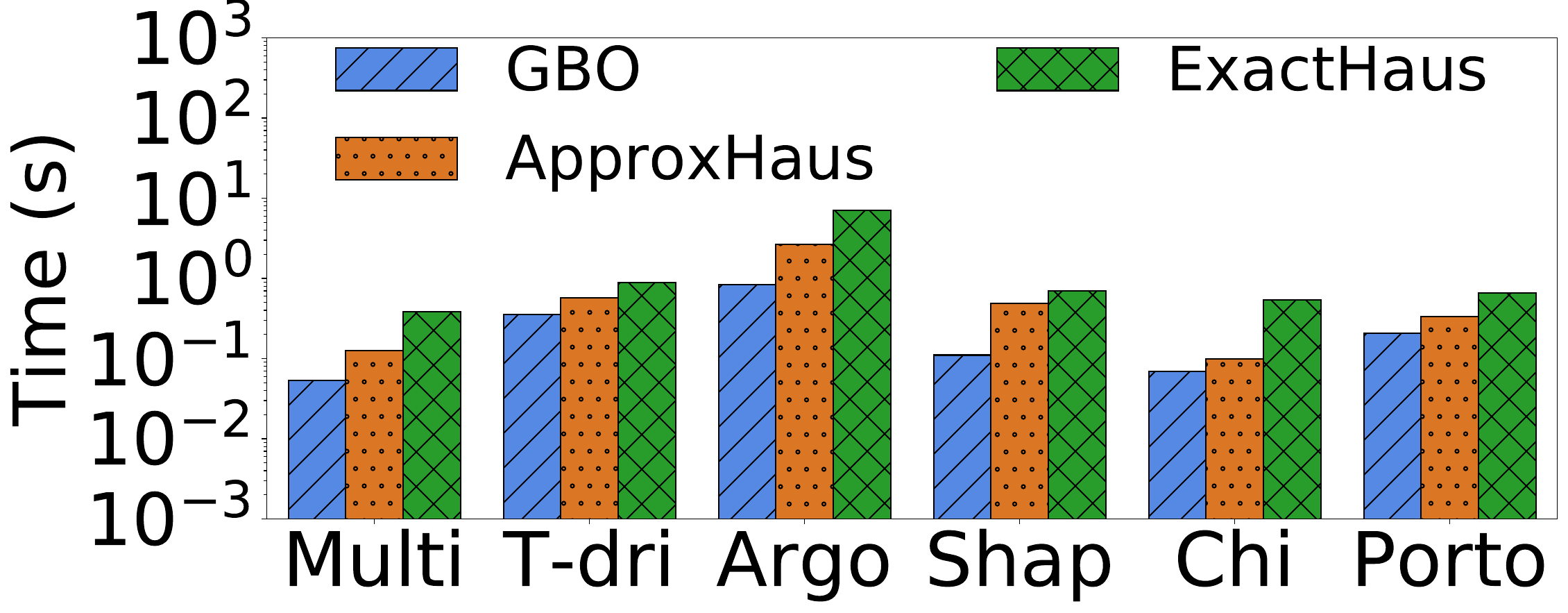}\vspace{-2.3cm}}
    \caption{Accuracy and runtime of three exemplar searches.}
\label{fig:resultHausExactAppro}
 \end{minipage}
 \begin{minipage}[b]{.49\textwidth}
  \centering
\subfigure[Accuracy of Outlier Removal]{\includegraphics[width=4.2cm]{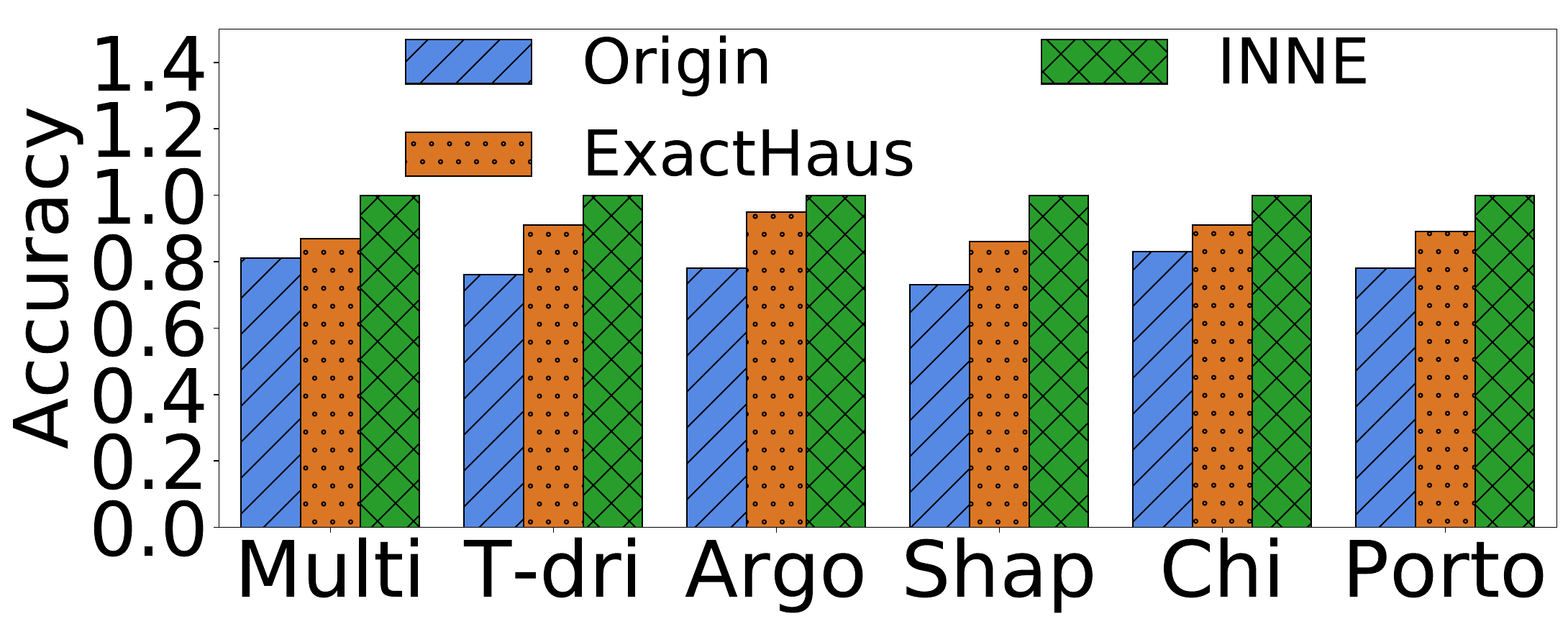}\vspace{-2.3cm}}
 \hspace{0cm}
 \subfigure[Runtime of Outlier Removal]{\includegraphics[width=4.2cm]{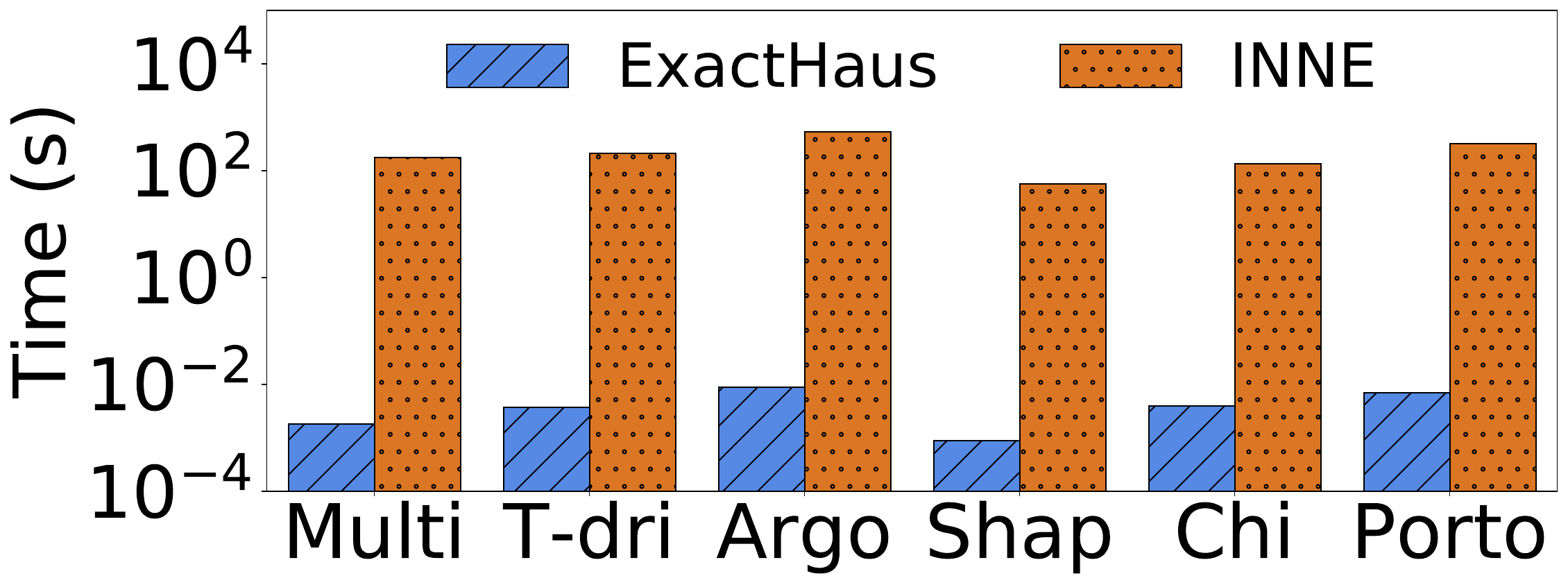}\vspace{-2.3cm}}
	\caption{Accuracy and runtime of top-$k$ Hausdorff search.}
		\label{fig:topkHausOutlier}
 \end{minipage}
 \vspace{-1.0em}
\end{figure*}

\begin{figure*}[t]
 \setlength{\abovecaptionskip}{0cm}
\setlength{\belowcaptionskip}{0cm}
 \begin{minipage}[b]{.49\textwidth}
  \centering
  \hspace{-1.5em}
\includegraphics[width=9cm]{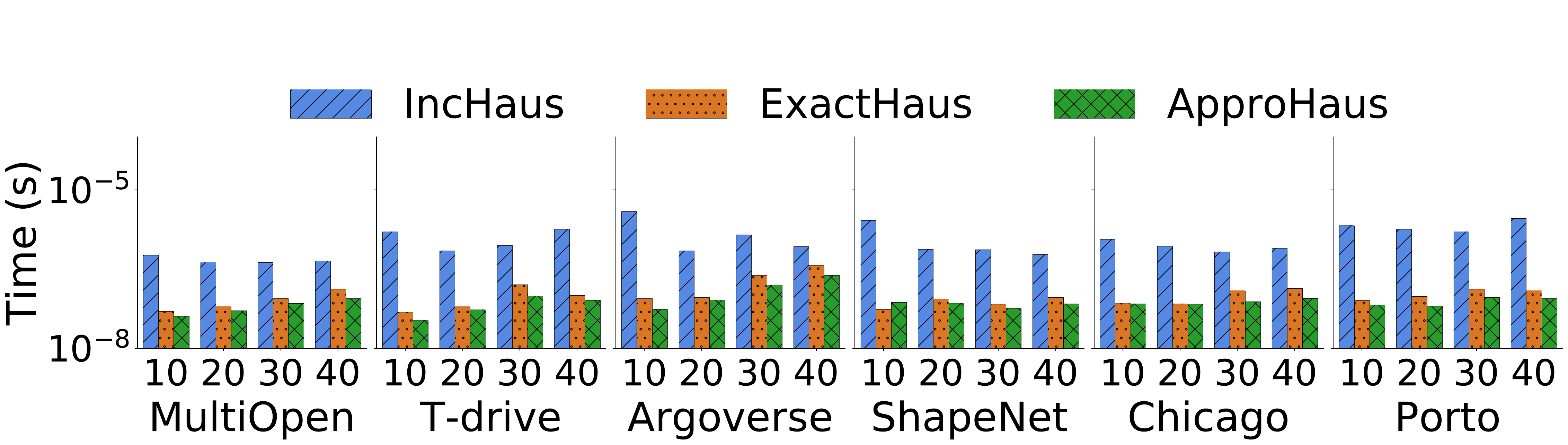}
    \caption{Capacity's effects on pairwise Hausdorff search.}
		\label{fig:capacityEffectPairwise}
 \end{minipage}
 \begin{minipage}[b]{.49\textwidth}
  \centering
 \includegraphics[width=9cm]{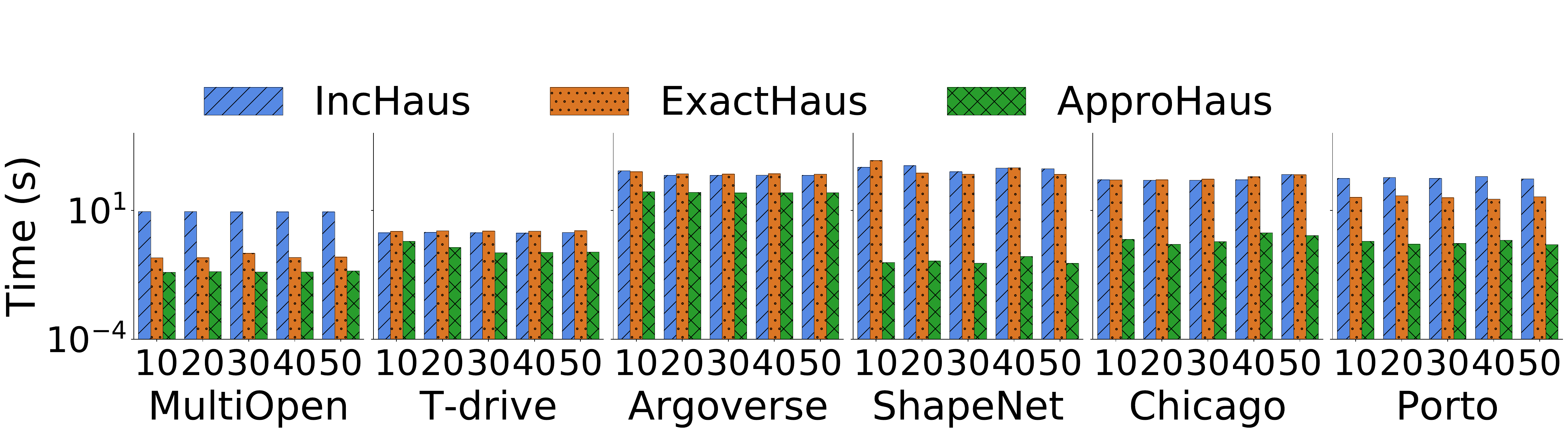}
    \caption{Index capacity's effects on top-$k$ Hausdorff search.}
		\label{fig:capacityEffect}
 \end{minipage}
 \vspace{-1.6em}
\end{figure*}

\begin{figure*}[t]
 \setlength{\abovecaptionskip}{0cm}
\setlength{\belowcaptionskip}{0cm}
 \begin{minipage}[b]{.49\textwidth}
  \centering
 	\subfigure[Runtime of overlap-based queries]{\includegraphics[width=4.2cm]{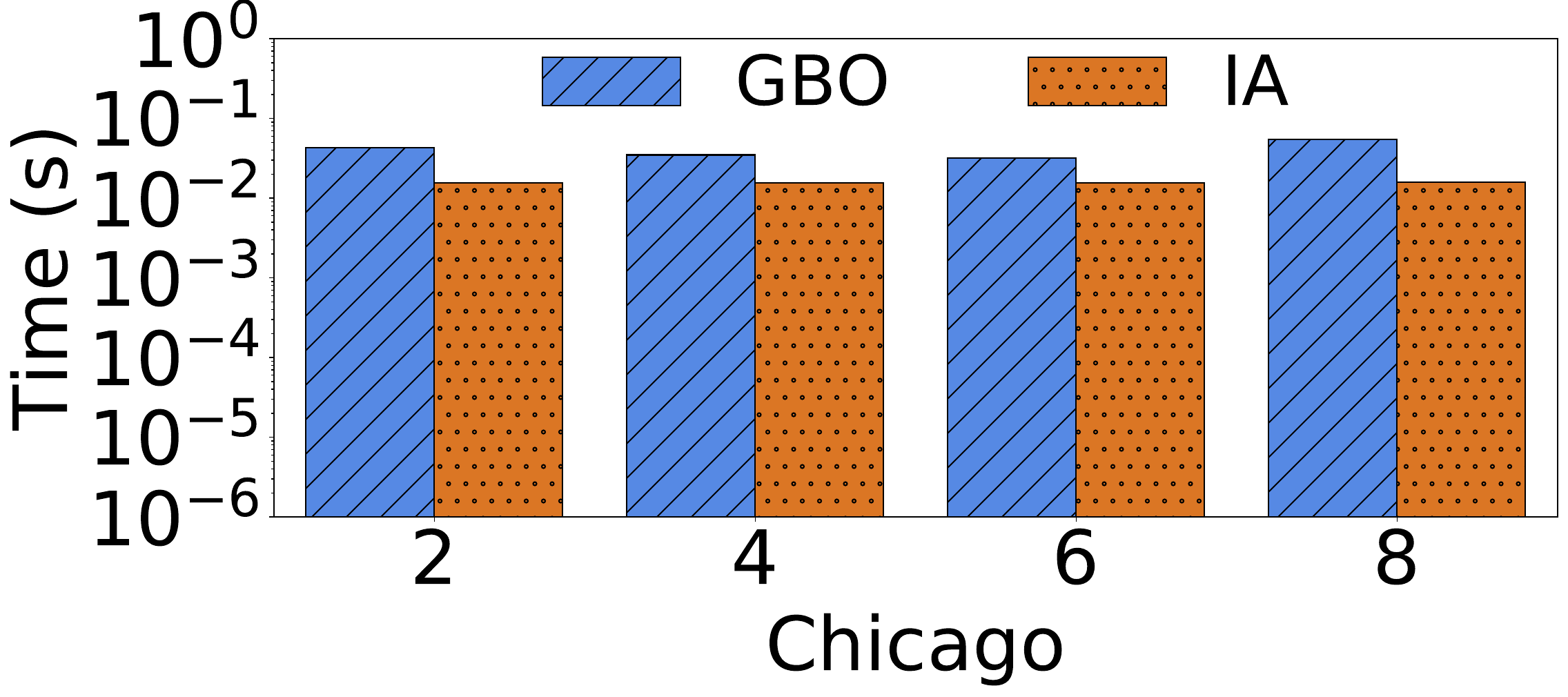}}
	\subfigure[Runtime of point-wise query]{\includegraphics[width=4.2cm]{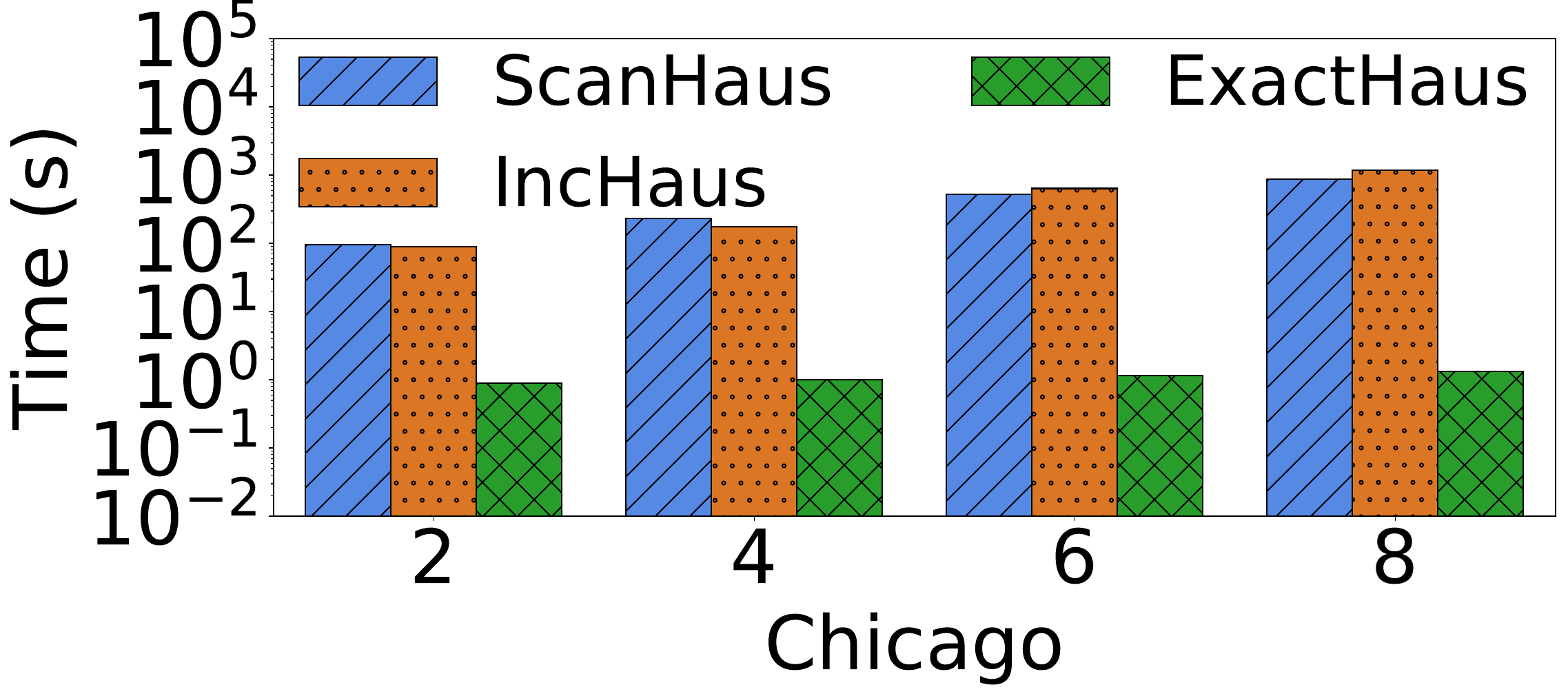}}
	\caption{Spatial dimension $d$'s effect on the Chicago dataset.}
    \label{fig:indexModedimensionEffect}
 \end{minipage}
 \hfill
 \hspace{0.2cm}
 \begin{minipage}[b]{.49\textwidth}
  \centering
 \includegraphics[width=8.5cm]{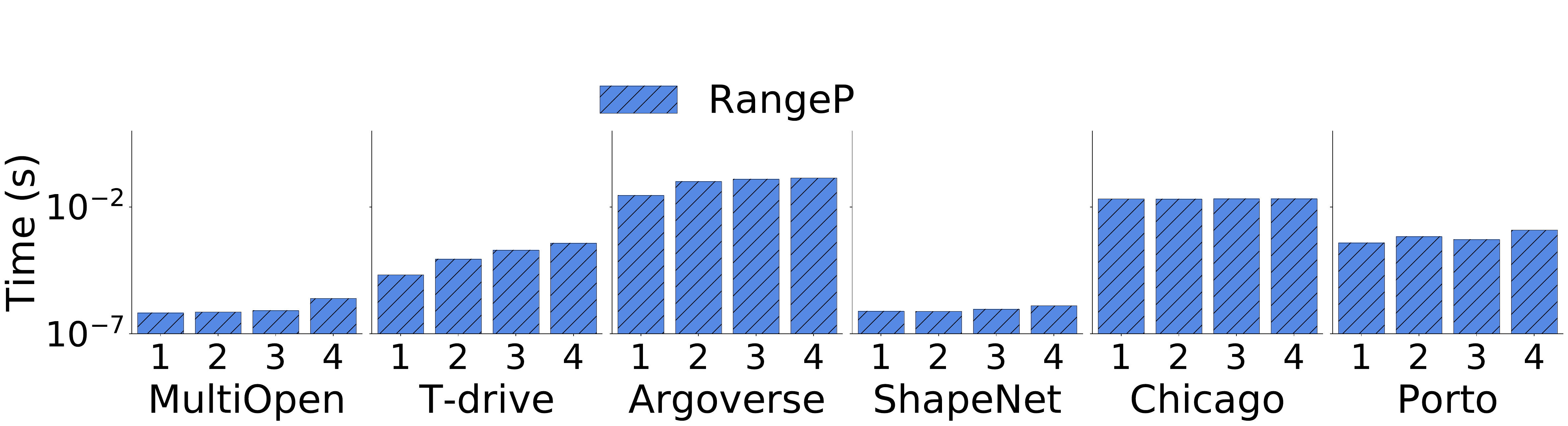}
	\caption{Range-based data point search as $R$ increases.} 
   \label{fig:joinUnionA}
 \end{minipage}
 \vspace{-1.7em}
\end{figure*}

\begin{figure*}[t]
 \setlength{\abovecaptionskip}{0cm}
\setlength{\belowcaptionskip}{0 cm}
 \begin{minipage}[b]{.49\textwidth}
  \centering
 	   \includegraphics[width=9cm]{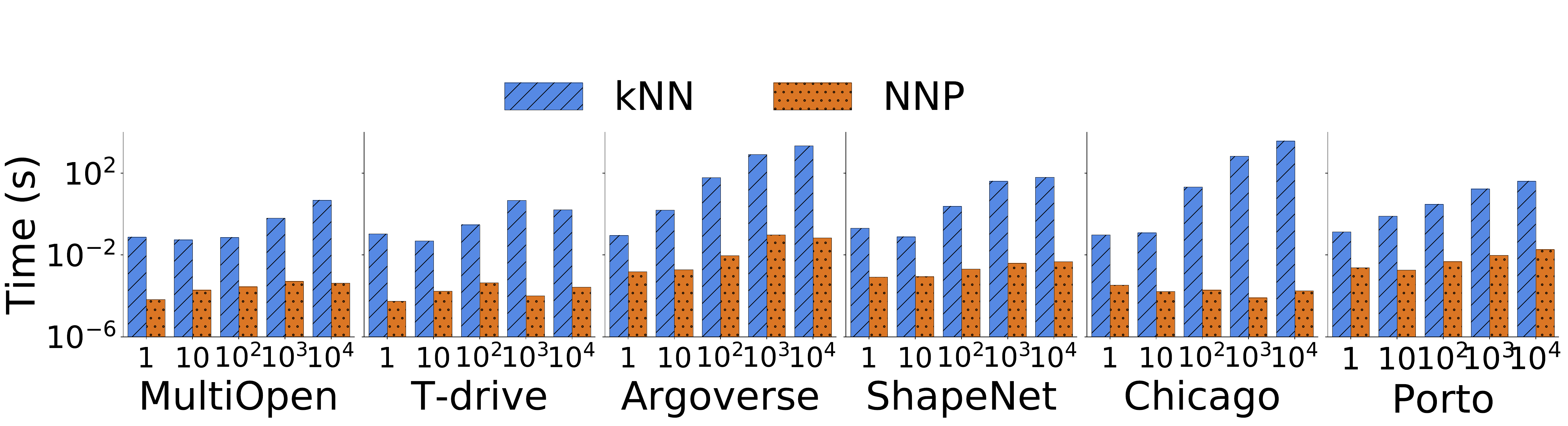}
	\caption{Nearest neighbor point search as $s$ increases.} 
\label{fig:joinUnionB}
 \end{minipage}
 \hfill
 \begin{minipage}[b]{.49\textwidth}
  \centering
\includegraphics[width=8.8cm]{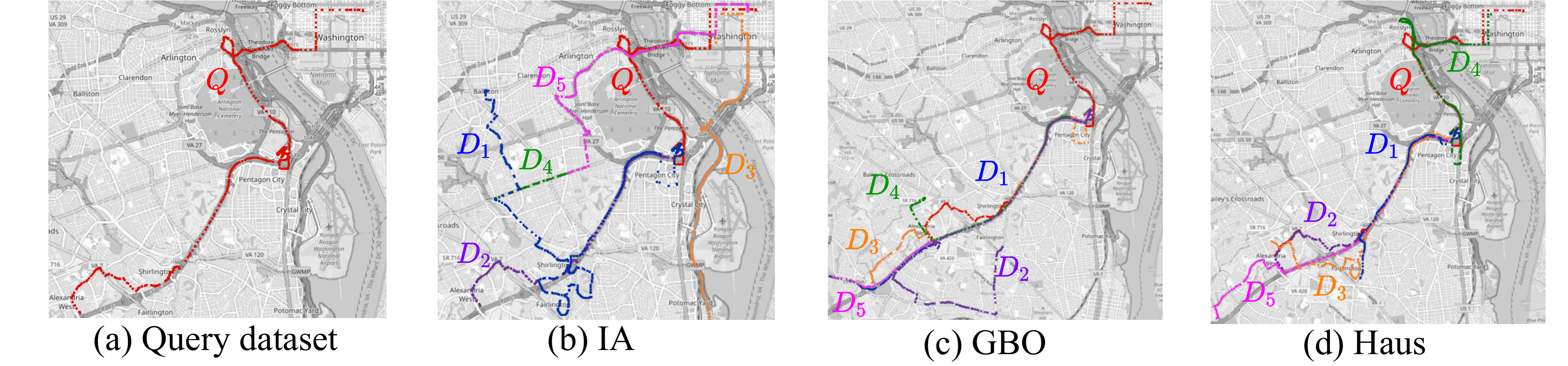}
	\caption{Case study on trajectory datasets.}
\label{fig:trajectory_case}
 \end{minipage}
 \vspace{-1.3em}
\end{figure*}

\myparagraph{Effect of Outlier Removal}
Figure~\ref{fig:topkHausOutlier} shows the experimental results of top-$k$ Hausdorff search under three conditions: without using outlier removal (\Origin), using our \ExactHaus, and utilizing the SOTA \INNE~\cite{bandaragoda2018isolation,zhao2019pyod} for outlier removal. The results from \INNE are set as the ground truth. Also, we compare the execution time of these two outlier removal algorithms. The results demonstrate that our outlier removal technique can effectively remove outliers in the dataset by detecting leaf nodes with excessively large radii, thereby enhancing the accuracy of the search (up to 90\%). Moreover, our algorithm exhibits significantly faster performance, surpassing \INNE by five orders of magnitude, due to the substantial reduction in data points achieved by detecting kneedle points.





\myparagraph{Effect of Node Capacity $f$}
 Figure~\ref{fig:capacityEffectPairwise} shows the runtime of the three index-based methods when computing the Hausdorff distance between a pair of datasets. We observe that as the $f$ increases, the execution time of the pairwise Hausdorff gradually increases. This is because larger leaf nodes imply that more data points need to be scanned one by one, which affects the performance. Figure~\ref{fig:capacityEffect} shows that as the capacity of leaf nodes $f$ increases, the running time of the three methods does not change much, indicating that the impact of node capacity on the top-$k$ queries is minimal. This is because more time is spent on pruning, sorting, and computing bounds, rather than scanning the data points within the nodes.
\myparagraph{Effect of Dimension $d$}
As the Chicago dataset's dimension is the highest, we use it to verify the performance of three representative operations in Figure~\ref{fig:indexModedimensionEffect}.
Firstly, from the left of Figure~\ref{fig:indexModedimensionEffect}, we can see that the query time of the two overlap-based methods does not change much with the increase of dimension, and \IA is much faster than \GBO. The right of Figure~\ref{fig:indexModedimensionEffect} shows the search time of the top-$k$ Hausdorff. The experimental result shows that with the increase of $d$, the running time of the \ScanHaus and \IncHaus increases, while the times of the \ExactHaus do not change significantly.
This indicates that dimension significantly affects the effectiveness of calculating distances between MBRs. In addition, to investigate the scalability of our proposed framework,  we also conduct experiments in Appendix~\ref{appendix:AdditionalResults}.

\vspace{-0.2cm}
\subsection{\textbf{Efficiency of Data Point Search}}
\label{exp:dataPointDiscovery}
\vspace{-0.1cm}
\myparagraph{Range-based Point Search}
Figure~\ref{fig:joinUnionA} shows the performance of searching all the points located inside a range. We can observe a slight increase in running time as the query range  $R$ increases. It is because as the query range increases, the number of candidates falling within the query range increases, so it takes more time to check the increasing number of candidates in a dataset.

\myparagraph{Nearest Neighbor Point Search}
We mainly compare our \NNP search operation presented in Section~\ref{sec:accDataPoint} against the k-nearest neighbor (\KNN)~\cite{Taha2015} that works similarly to the Hausdorff distance \cite{Nutanong2011}. Figure~\ref{fig:joinUnionB} shows that our method can accelerate the \NNP operation significantly, which achieves up to seven orders of magnitude speedup over \KNN. This is because the \KNN shares a common procedure with \NNP to find the nearest points for points in the query datasets. By continuously traversing the unified index, our method reduces computation time substantially.

\subsection{Case Study}
\label{caseStudy}
Figure~\ref{fig:trajectory_case} shows the visualization of the results that conduct trajectory search and analysis based on the \spadas system developed in this paper. Specifically, we added a transportation dataset collection to \spadas that includes a variety of trajectory datasets from Washington, D.C.
An analyst aims to find $k$ trajectory datasets that are closely similar to a given query dataset from the \spadas to perform trajectory analysis tasks, such as trajectory near-duplicate detection~\cite{XiaoWLYW11} and traffic congestion prediction~\cite{ZhengZYSZ14,ZhouTWN13}.


As shown in Figure~\ref{fig:trajectory_case}(a), the analyst first inputs a query dataset $Q$ from Washington, D.C. into \spadas, aiming to find the top 5 datasets that are similar to $Q$. The results found based on \IA, \GBO, and \Haus similarity measures are shown in Figures~\ref{fig:trajectory_case}(b), (c), and (d). We can see that the result datasets obtained by \IA are coarse-grained compared with \GBO and \Haus. For example, although the intersecting area between the result dataset $D_5$ and $Q$ is quite large, their data point distributions are not similar.
In contrast, the results obtained by \Haus and \GBO are more similar to the query $Q$, but \Haus offers a more fine-grained similarity. For example, the result $D_2$ obtained by \Haus is more similar to $Q$ in terms of dataset distribution than $D_2$ obtained by \GBO. The analyst can use these retrieved datasets, along with $Q$, for trajectory data analysis tasks.



\section{Conclusions}
\label{sec:con}
We presented \spadas, a multi-granularity spatial data search system that seamlessly supports coarse-grained dataset search based on multiple distance metrics and fine-grained data point search.  We first designed a unified two-level index, which can effectively organize spatial datasets and data points to support multiple types of search operations, and mitigate the impact of outliers on the accuracy of similarity measures. Then, we proposed a range of pruning mechanisms, including fast bound estimation, approximation technique with error bound, and pruning in batch techniques.
Finally, we conducted extensive experiments using six real-world data repositories. The results corroborated that the proposed system is effective, efficient, and scalable. In future work, we plan to further evaluate the effectiveness of additional distance measures.


\balance
\bibliographystyle{abbrv}
\bibliography{url}

\preto{\section}{\setcounter{cN}{0}} 

\clearpage
\section{Appendix}
\label{Appendix1}

\subsection{Notations}
\label{appendix:notations}
Frequently used notations are summarized in Table~\ref{tab:notations}.

\begin{table}[h]
	\centering
 	\caption{Summary of notations.}
  \label{tab:notations}
\renewcommand\arraystretch{1}\addtolength{\tabcolsep}{1.0pt}
		\begin{tabular}{cc}
			\toprule
			\textbf{Notation}	& \textbf{Description} \\ \midrule
			$p$, $D$, $\mathcal{D}$	& a data point, a dataset and a data repository \\ 
   $d$ & the dimension of spatial data point\\
			$|D|$	& the number of spatial points in the dataset $|D|$ \\ 
			$|\mathcal{D}|$ &  the number of datasets in the data repository $|\mathcal{D}|$ \\
			$R$, $Q$	& a query range  and a query dataset \\
    $b^\downarrow, b^\uparrow$ & \makecell{the bottom left corner and upper right corner of the \\ minimum bounding rectangle (MBR) of the dataset} \\
			$2^\theta, \theta$ & the number of cells in each dimension, resolution\\
     $z(D)$ & the $z$-order signature of dataset $D$\\
     $N_D$ & a dataset root node corresponding to the dataset $D$ \\ 
     $o$, $r$ &\makecell{center point and radius of index node}\\
     $N_l$, $N_r$ &\makecell{left and right child nodes of index node}\\
     $f$ & the leaf nodes' capacity \\
     $\phi$ & a sorted list to store all leaf nodes' radii \\
			$\epsilon$ & the threshold parameter for approximate Hausdorff \\

			\bottomrule
		\end{tabular}
\end{table}


\subsection{Time Complexity Analysis}
\label{appendix:indexComplexity}
We use $n$ to represent the number of points in each dataset, and $m$ to represent the number of datasets in the data repository. The construction of the unified index involves both the construction of the bottom-level index and the upper-level index. In constructing the bottom-level index for each dataset. Algorithm~1 first splits the data points into two child sets from which trees are recursively built. During each splitting step, the time complexity for splitting all data points is $O(n)$, and the height of the bottom-level index is $\log(n)$. Thus, the time complexity is $O(n\log n)$. 

For outlier removal, the bottom-level index needs to sort the radii of all leaf nodes and compute the radius threshold $r'$, resulting in a time complexity of $O(n\log n)$.  Finally, the bottom-level index traverses up from all leaf nodes to remove those potential outliers and refine the dataset index node, resulting in a time complexity of $O(2n)$.  Thus, the total time complexity of the bottom-level index for each dataset is $O(n\log n + n\log n + 2n ) = O(n\log n)$.

In contrast, the construction process of the upper-level follows a similar approach to that of the bottom-level index, yielding a time complexity of $O(m \log m)$. Since the number of datasets $m$ is significantly smaller than the number of data points $n$ in each dataset $D$, $m$ can be considered negligible in comparison to $n$.
Thus, the total time complexity of constructing the unified index is $O(m \log m + m(O(n\log n))) = O(mn \log n)$.

\subsection{Additional Experimental Results}
\label{appendix:AdditionalResults}
\myparagraph{Effect of High-dimensional Dataset} 
To investigate the scalability of our proposed index, we also conduct experiments to demonstrate the construction cost and search performance of our proposed unified index and algorithm on high-dimensional datasets. Here, we utilize the highD dataset~\cite{highDdataset}, a large-scale 25-dimensional user trajectories dataset collected from German highways. The highD dataset is publicly available online\footnote{The highD dataset is available online: \url{https://levelxdata.com/highd-dataset/}}. Additionally, comprehensive details on the dataset format and column descriptions are provided\footnote{The dataset format information can be found at: \url{https://levelxdata.com/wp-content/uploads/2023/10/highD-Format.pdf.}}.

Specifically, we first compare the construction cost on the highD data repository between our unified index and the independent index~\cite{Nutanong2011} of \IncHaus. Figure~\ref{fig:highDConstructionCost} shows the index construction time comparison of two indexes as the data repository scale $m$ increases. We can see that the construction time for both indexes rises as the data repository scale $m$ increases, which is expected, as larger $m$ imposes greater computational burdens, further increasing the construction time.  

In addition, combined with Figure~\ref{fig:indexCompare} in Section~\ref{exp:index}, we can observe that the construction time for both indexes is higher on the high-dimensional data repository compared to the low-dimensional one. This is because the increase in dimension leads to a decrease in the efficiency of space partitioning when building a tree. However, our proposed unified index still consistently outperforms \IncHaus, which achieves at most 23.3\% speedup over the index of \IncHaus.


Subsequently, we compare the search performance of our \ExactHaus and \ApproHaus with \IncHaus and \ScanHaus on the highD data repository as the $k$ increases. The search experimental results are shown in Fig~\ref{fig:highDIncreaseK}. With the increase of $k$, we can observe that \ExactHaus and \ApproHaus drastically improve the search performance of the top-k Huasdorff search compared with the other two comparison algorithms, achieving speed improvements of up to several orders of magnitude. This indicates that our algorithms based on the unified index can effectively filter out unpromising datasets even in high-dimensional scenarios. In addition, we can see that the \ApproHaus is very fast, completing the search process within one second, which provides faster response time in high-dimensional datasets. Overall, these experimental results verify that our index and search algorithms are very scalable on the high-dimensional spatial data repository.




\myparagraph{Comparison of Index and Search Algorithms}
To further investigate the efficiency of our proposed framework and algorithms on a specific data repository at a more granular level, we added an experiment to compare the time cost of constructing the integrated index and dataset search and data point search in the \textbf{Chicago} data repository. Firstly, we present the experimental results of the dataset search using multiple similarity metrics, including \RangeS, \IA, \GBO, and \ExactHaus. In contrast, in the data point search, we provided the experimental results of \RangeP and \NNP.

From Figure~\ref{fig:indexCompare} we can observe that as the data repository scale $m$ increases, the running time of index construction and dataset search algorithm gradually increases, while the running time of data point search does not significantly change. This is because the data point search defined in Section~\ref{sec:enrich} is between the query dataset and the specified result dataset, which is not affected by the scale of the data repository.

Additionally, we can also see that with the increase of $m$, the running time of \ExactHaus search is longer compared to the other search methods. This is due to the fact that the Hausdroff distance requires the computation of pairwise point distance, which consumes a lot of time. However, the running time still remains under one second, demonstrating the effectiveness of the index-based acceleration algorithm we have designed.

\begin{figure}[t]
	\centering
	\hspace{-1em}\includegraphics[width=7.8cm]{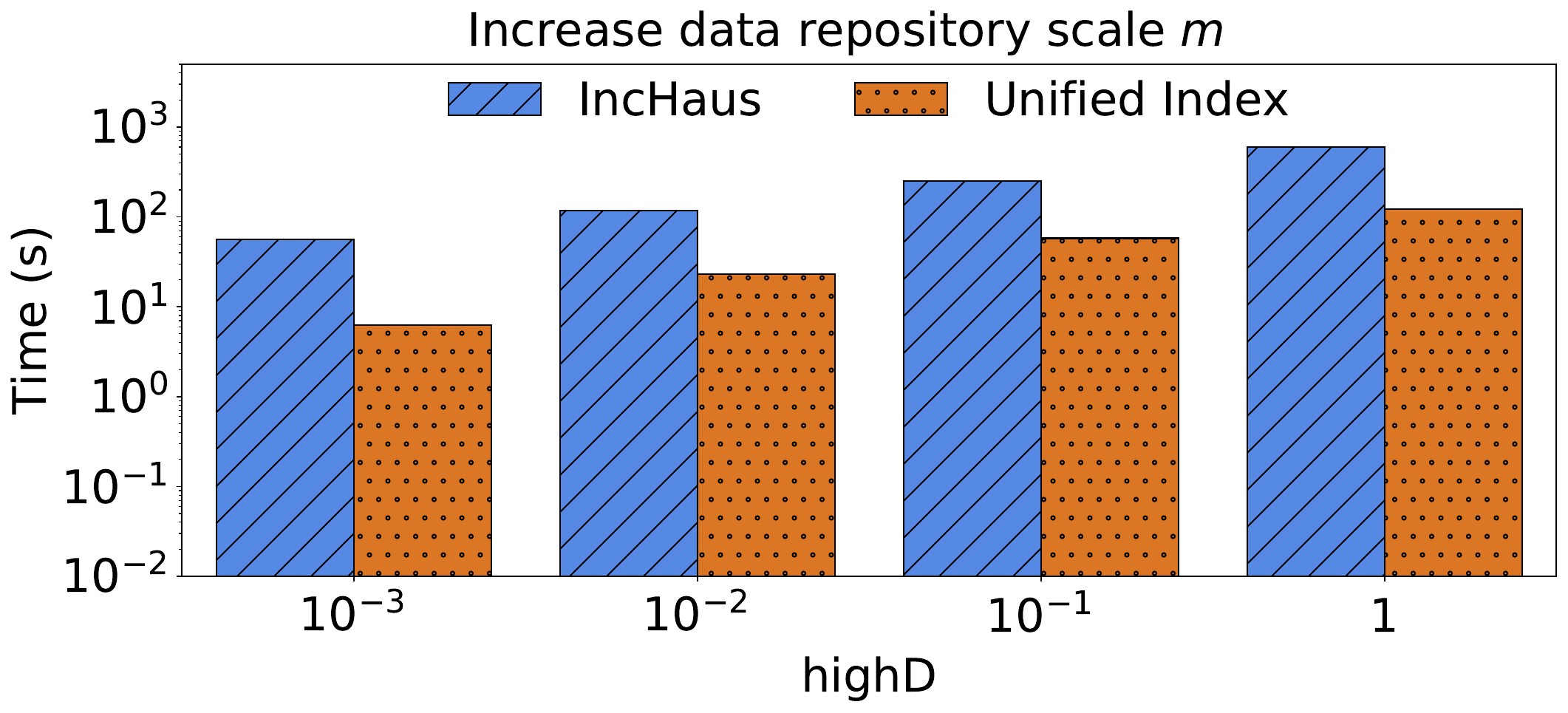}
	\vspace{-1.2em}
	\caption{Index's construction time with increasing data repository scale $m=\{10^{-3}, 10^{-2}, 10^{-1}$, 1\}$\times|\mathcal{D}|$.}
	\label{fig:highDConstructionCost}
\end{figure}

\begin{figure}[t]
	\centering
      \setlength{\abovecaptionskip}{0cm}
\setlength{\belowcaptionskip}{0cm}
\hspace{-1em}\includegraphics[width=7.8cm]{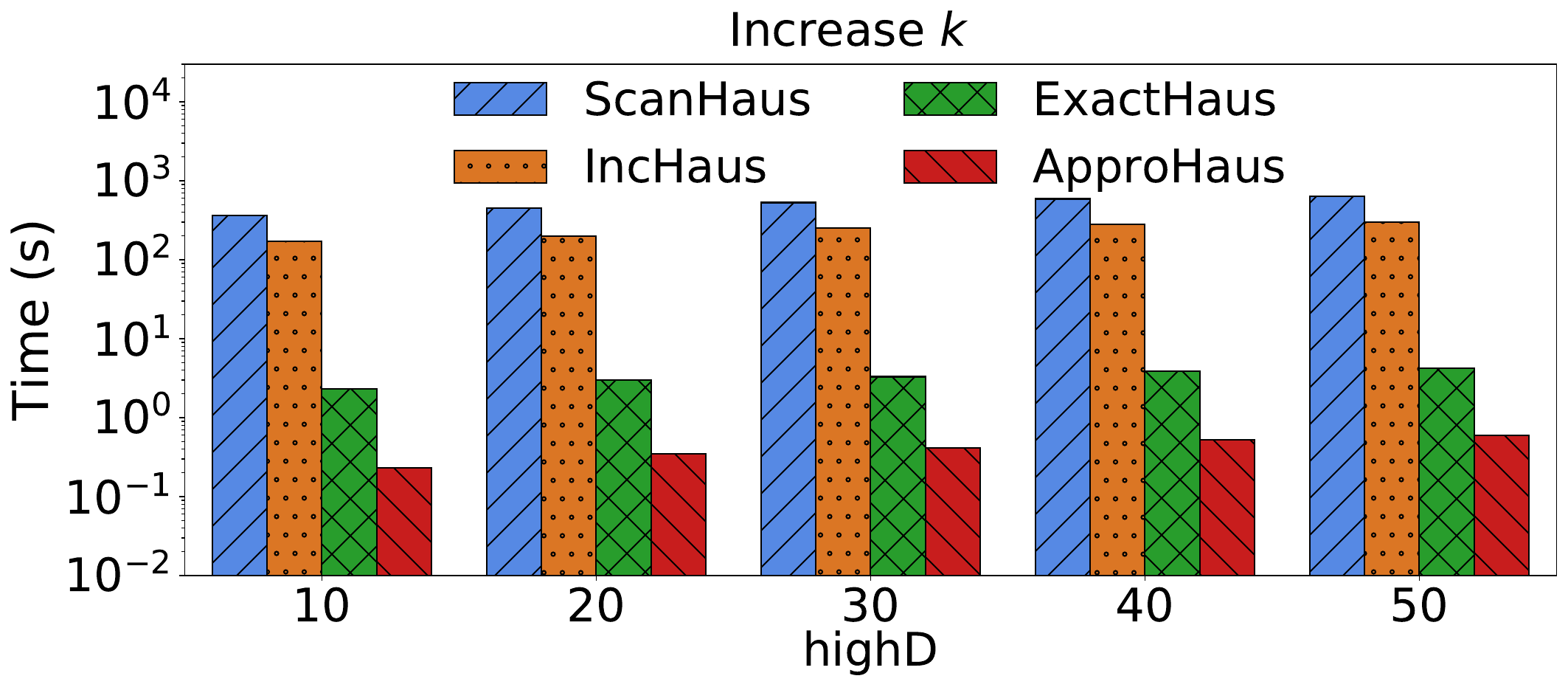}
	\caption{Top-k Hausdorff search with increasing $k$.}
	\label{fig:highDIncreaseK}
\end{figure}

\begin{figure}[t]
	\centering
\setlength{\abovecaptionskip}{0cm}
\setlength{\belowcaptionskip}{0cm}
\hspace{-1em}\includegraphics[width=9cm]{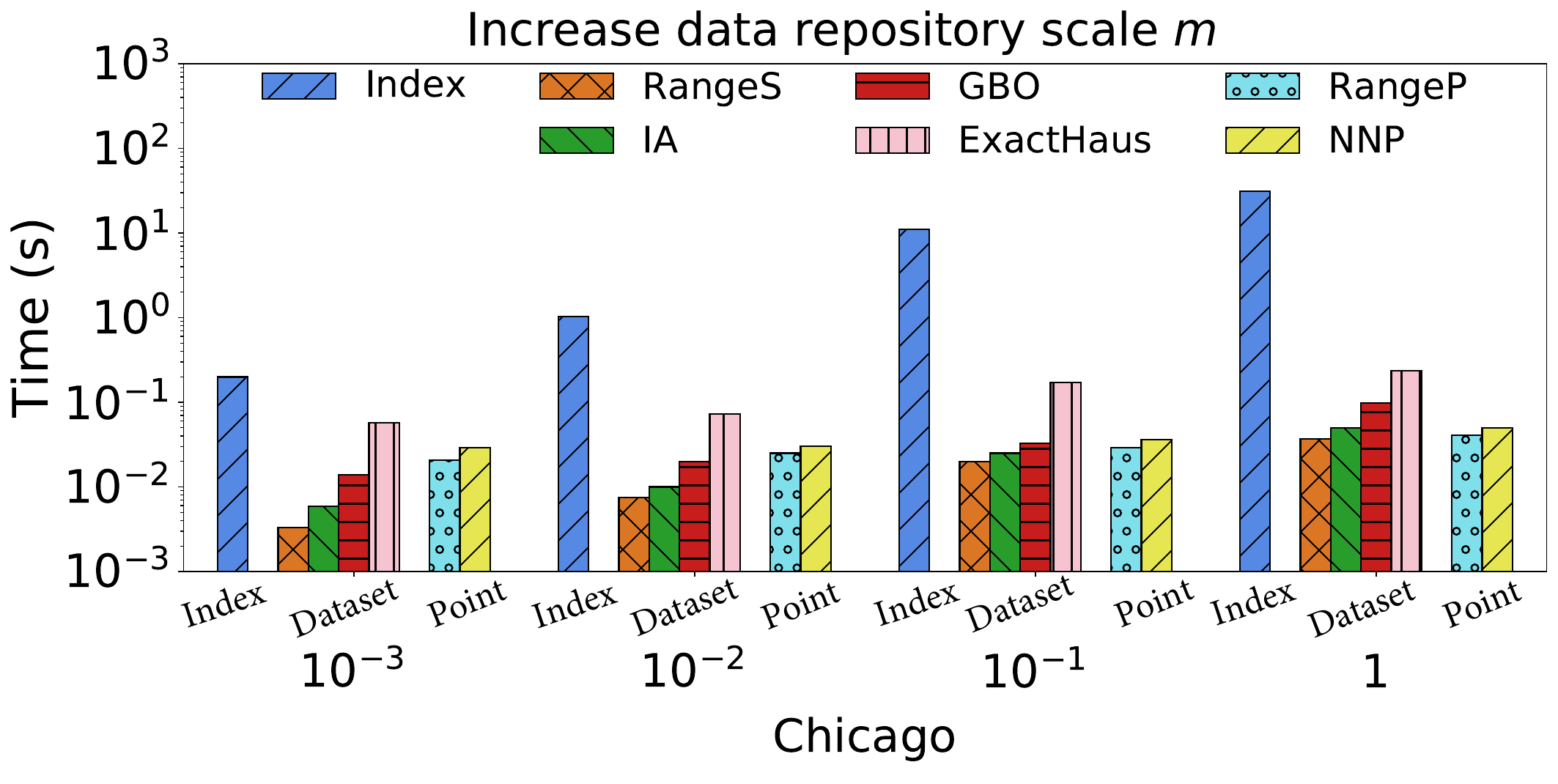}
	\caption{Index's construction time and the search time of dataset search and data point search with increasing scale $m$ in \textbf{Chicago} data repository.}
\label{fig:indexCompare}
\end{figure}

\end{document}